\documentclass[sn-mathphys,Numbered]{sn-jnl}% Math and Physical Sciences Reference Style
\usepackage{graphicx}%
\usepackage{multirow}%
\usepackage{amsmath,amssymb,amsfonts}%
\usepackage{amsthm}%
\usepackage{mathrsfs}%
\usepackage[title]{appendix}%
\usepackage{xcolor}%
\usepackage{textcomp}%
\usepackage{manyfoot}%
\usepackage{booktabs}%
\usepackage{algorithm}%
\usepackage{algorithmicx}%
\usepackage{algpseudocode}%
\usepackage{listings}%
\usepackage{ulem}

\theoremstyle{thmstyleone}%
\theoremstyle{thmstyletwo}%

\theoremstyle{thmstylethree}%

\renewcommand{\vec}[1]{\mbox{\boldmath $#1$}}

\begin{document}

\title[Article Title]{Dineutron clusters}

%%=============================================================%%
%% Prefix	-> \pfx{Dr}
%% GivenName	-> \fnm{Joergen W.}
%% Particle	-> \spfx{van der} -> surname prefix
%% FamilyName	-> \sur{Ploeg}
%% Suffix	-> \sfx{IV}
%% NatureName	-> \tanm{Poet Laureate} -> Title after name
%% Degrees	-> \dgr{MSc, PhD}
%% \author*[1,2]{\pfx{Dr} \fnm{Joergen W.} \spfx{van der} \sur{Ploeg} \sfx{IV} \tanm{Poet Laureate} 
%%                 \dgr{MSc, PhD}}\email{iauthor@gmail.com}
%%=============================================================%%

\author*[1]{\fnm{Takashi} \sur{Nakamura}}\email{nakamura@phys.sci.isct.ac.jp}

\author[2,3,4]{\fnm{Kouichi} \sur{Hagino}}\email{hagino.kouichi.5m@kyoto-u.ac.jp}
\equalcont{Kouichi Hagino and Yosuke Kondo contributed equally to this work.}

\author[4,1]{\fnm{Yosuke} \sur{Kondo}}\email{kondo@ribf.riken.jp}
\equalcont{Kouichi Hagino and Yosuke Kondo contributed equally to this work.}

\affil*[1]{\orgdiv{Department of Physics}, \orgname{Institute of Science Tokyo}, 
\orgaddress{\street{2-12-1 O-Okayama}, \city{Meguro}, \state{Tokyo}, \postcode{152-8550},  \country{Japan}}}

\affil[2]{\orgdiv{Department of Physics, Graduate School of Science}, \orgname{Kyoto University}\orgaddress{\street{}, \city{Kyoto}, \postcode{606-8502}, 
%\state{Kyoto}, 
\country{Japan}}}

\affil[3]{\orgdiv{Institute for Liberal Arts and Sciences}, \orgname{Kyoto University}\orgaddress{\street{}, \city{Kyoto}, \postcode{606-8501}, 
%\state{Kyoto}, 
\country{Japan}}}

\affil[4]{\orgname{RIKEN Nishina Center}, \orgaddress{\street{Hirosawa 2-1}, \city{Wako}, \state{Saitama}, \postcode{351-0198}, \country{Japan}}}

%%==================================%%
%% sample for unstructured abstract %%
%%==================================%%

\abstract{The dineutron is a spatially compact two-neutron cluster, which is expected to appear in a low-density part of nuclei.  In recent years, there has been rapid progress in experimental and theoretical research on dineutron clusters, particularly on neutron-rich rare isotopes.  Experimentally, evidence for dineutron in two-neutron halo nuclei, such as $^{11}$Li, has been obtained using Coulomb breakup, measurements of charge radii, and quasi-free proton scattering. Specific unbound nuclei just beyond the neutron drip line, which decay by emitting two neutrons, are also candidates for having a dineutron correlation. For instance, the dineutron structure has recently been investigated for $^{16}$Be, focusing on its decay into the core and the two neutrons.  
Theoretically, it is shown that the dineutron is partially due to the admixture of different-parity configurations for the two valence neutrons. Few-body theories, including dynamical effects of the decay process, play important roles in interpreting three-body decays. 
We also discuss the four-neutron clusters, showing the experimental results of recent tetraneutron experiments and observation of $^{28}$O. Possible relevance of these states to dineutron correlation is discussed.
Finally, we discuss future perspectives on dineutron clusters in neutron-rich nuclei and their relation to the universal features in few-body physics. }


%%\pacs[JEL Classification]{D8, H51}

%%\pacs[MSC Classification]{35A01, 65L10, 65L12, 65L20, 65L70}

\maketitle

\section{Introduction}\label{sec:intro}

Pairing is one of the most fundamental correlations in atomic nuclei~\cite{BRIBRO05,RingSchuck}. 
The pairing correlation is included in the Weizs\"{a}cker-Bethe's nuclear mass formula: even-even nuclei are more stable than odd-even and odd-odd nuclei due to the
extra stability of two neutrons (protons) with anti-parallel spins ($S=0$). 
The pairing correlation is attributed to the BCS (Bardeen-Cooper-Schrieffer) mechanism, as in superconductors.
 BCS pairing is known to exhibit long-range behavior, with the coherence length exceeding the nuclear diameter~\cite{BRIBRO05}.
On the other hand, Migdal showed that two neutrons may form a quasi-bound dineutron on the surface of a nucleus, which can be spatially compact,
and, as such, exhibits very different behavior compared to the BCS pairing~\cite{MIGD73}.
Experimental evidence for such a dineutron has only recently been obtained for two-neutron halo nuclei, such as $^{11}$Li~\cite{NAKA06,KUBO20}, 
and unbound resonances such as $^{16}$Be~\cite{MONT24}, 
which will be discussed here.
 
\begin{figure}[ht]
\centering
\includegraphics[width=100.mm]{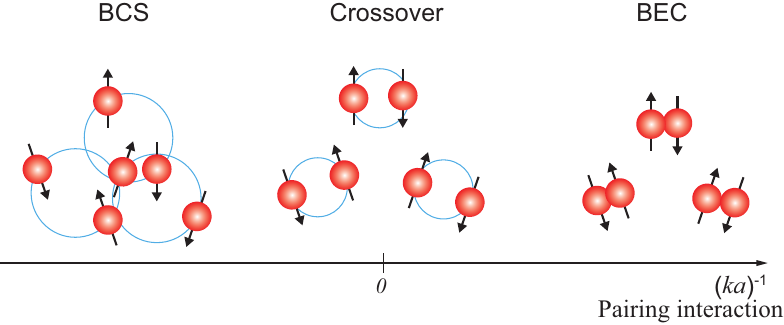}
\caption{Transition of a pair of fermions from the BCS to BEC regime is shown schematically as a function of pairing interactions, $(ka)^{-1}$, where $k$
is the Fermi momentum and $a$ is the scattering length. From the BCS regime to the BEC regime, the interaction becomes stronger. The intermediate region
is called the crossover  
}\label{fig:BCSBEC}
\end{figure}

 Figure \ref{fig:BCSBEC} shows the transition of the correlations for a pair of  fermions at temperature $T=0$
 from the BCS regime to the Bose-Einstein Condensate (BEC) regime as a function of the interaction between them.
 The intermediate regime is called the BCS-BEC crossover. 
 The interaction is characterized by $(ka)^{-1}$, where $a$ is the $s$-wave scattering length between the two particles, and $k$ is the Fermi momentum
 of the particles.  The crossover regime appears in $-1 <  (ka)^{-1} <1$, and the boundary between the positive and negative $a$ values, where $(ka)^{-1}=0$ ($a\rightarrow \infty$), is called the unitary limit.
 The BCS regime corresponds to weak coupling, where the pairing interactions are weaker, while the BEC regime corresponds to strong coupling, with stronger interactions.
Note that this picture is universal for fermionic systems at all scales, from molecules and atoms to nuclei and 
hadrons.~\cite {NAKA25, HORI25}.

\begin{figure}[ht]
\centering
\includegraphics[width=90.mm]{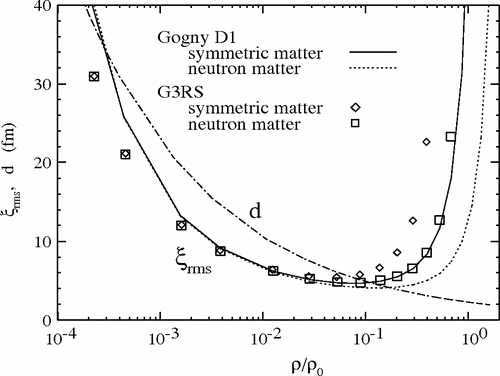}
\caption{ Rms radii of the neutron cooper pair, $\xi_{rms}$, calculated for the Gogny D1 force (solid line: symmetric matter, dotted line: neutron matter) and the G3RS force (diamond symbols: symmetric matter, square symbols: neutron matter), are compared with the average interneutron distance $d(=\rho^{-1/3})$ (dot dashed line) as a function of the neutron density $\rho/\rho_0$. The dineutron correlation region corresponds to $d>\xi_{rms}$, which appears for $10^{-4} <\rho/\rho_0 < 0.1$. The figure is adopted from Ref.\cite{MATS06}
}\label{fig:densedep}
\end{figure}
 
 For a pair of neutrons in nuclei, the interactions can be varied according to $1/k$, while $a$ for the $s$-wave is constant: $-$18.9(4)~fm~\cite{MACH01, GARD09}.
 Note that the value $|a|$ for $nn$ is sufficiently large compared to the range of the neutron-neutron interaction. As such, a pair of neutrons should have a property arising from proximity to the unitary limit.
 Since $k$ is a function of the density, $k= (3\pi^2 \rho)^{1/3}$, as approximated in the Fermi-gas model,  the $nn$ interaction becomes stronger for a lower density.  
 The transition of the neutron pairing correlation as a function of the nuclear matter density was discussed by Matsuo~\cite{MATS06}. 
 As shown in Fig.~\ref{fig:densedep}, the root-mean-square (rms) radius of the neutron Cooper pair $\xi_{rms}$ 
 (correlated $S=0$ pair) and the mean distance $d~(=\rho^{-1/3})$ are plotted as a function of $\rho/\rho_0$, where $\rho$ is the nuclear matter density 
 and $\rho_0$ is the
 nuclear saturation density. For both neutron matter and symmetric nuclear matter, irrespective of the interactions adopted (G3RS: bare nucleon-nucleon interactions, Gogny-D1: effective
 nucleon-nucleon interactions with D1 parameters), it was shown that $\xi_{rms}$ becomes smaller than $d$ for $10^{-4}<\rho/\rho_0<0.1$, indicating that spatially 
 compact $nn$ correlations appear in such a low-density nuclear environment. Ref.~\cite{MATS06} 
 also demonstrated that at these densities, a pair of neutrons can be in the BCS-BEC crossover region. 
 We refer to such a spatially-correlated, compact $nn$ system as a ``dineutron" here, which is in line with the concept of a quasi-bound dineutron introduced by Migdal. In Ref.~\cite{MATS06}, strong spatial correlations at short distances are also shown at a moderate density of $\rho/\rho_0\sim 0.5$, indicating that dineutron correlations are significant over the density range, $10^{-4}<\rho/\rho_0<0.5$, in uniform superfluid nuclear matter.
 
 Two-neutron halo nuclei, 
 which have a peculiar three-body structure composed of a saturated core and two extended valence neutrons,
 may have dineutron correlation, as the density of the halo is much lower than the saturation density. Indeed, experimental evidence for dineutron has been accumulated 
 for two-neutron halo nuclei: $^{6}$He~\cite{SUN21,MUEL07}, $^{11}$Li~\cite{NAKA06,KUBO20}, $^{14}$Be~\cite{CORS23}, $^{17}$B~\cite{CORS23}, and $^{19}$B~\cite{COOK20}. 
 The dineutron correlation can be studied using Coulomb breakup,
 charge-radii measurements, 
 and quasi-free $(p,pn)$ reactions. 
 
 The other candidates for the dineutron correlation are unbound two-neutron emitters, such as $^{16}$Be~\cite{SPYR12,MONT24} and $^{26}$O~\cite{HAGI14,HAGI16}. 
 The excited states of neutron-rich nuclei, such as
 $^{6}$He($2_1^+$)~\cite{KIKU13}, may also have dineutron correlation. In these cases, the three-body decay into the core $+n+n$ can be observed, and its characteristic decay-correlation properties may provide a signal for the dineutron structure.
 
 Note that there has been confusion regarding the concept of the dineutron.
 Two neutrons with very low relative energies in a decay can be correlated due to final-state interactions (FSI) between them. Some regard this phenomenon as dineutron decay, 
 whereas here we do not treat it as a dineutron correlation, but rather as
 $nn$ correlations arising from FSI. In this article, we define a spatially correlated $nn$ system as a dineutron. We discuss this issue in the case of $^{16}$Be.
Theoretical descriptions of the three-body decay of dineutron-correlated states play an important role in distinguishing between these FSI-induced $nn$ correlations and dineutron correlations in the nucleus. 
 
This article is organized as follows:  
Section~\ref{sec:di} discusses the dineutron correlation for two-neutron halo nuclei.
Section~\ref{sec:diexp} reviews recent experimental results for
dineutrons in halo nuclei. 
Then in Section~\ref{sec:ditheo}, the theoretical 
aspects of dineutrons in halo nuclei are discussed.
Section~\ref{sec:unbound} discusses the dineutron correlation in unbound nuclei (two-neutron emitters), where the experimental and theoretical aspects are
shown in Sections~\ref{sec:unboundexp} and \ref{sec:unboundtheo}, respectively.
Section~\ref{sec:tetra} describes the experimental results on tetraneutron~\cite{KISA16,DUER22} and $^{28}$O~\cite{KOND23}. The latter is a doubly-magic nucleus candidate and one of the very rare four-neutron emitters. 
We discuss the dineutron-dineutron correlations of this nucleus.
We also briefly discuss the dineutron and $4n$ correlations in $^{8}$He.
Finally, in Section~\ref{sec:summary}, the summary and the near-future perspectives are presented.

\newcommand{\erel}{E_{\rm rel}}
\newcommand{\ex}{E_{\rm x}}
\newcommand{\eone}{E{\rm 1}}
\newcommand{\beone}{B(\eone)}
\newcommand{\dsderel}{\frac{d\sigma_{\rm CB}}{dE_{\rm rel}}}
\newcommand{\neone}{N_{\eone}(E_{\rm x})}
\newcommand{\sn}{S_{n}}
\newcommand{\snn}{S_{2n}}
\newcommand{\dbderel}{\frac{dB(\eone)}{dE_{\rm rel}}}
\newcommand{\dsderelone}{d\sigma_{\rm CB}/dE_{\rm rel}}
\newcommand{\dbderelone}{dB(\eone)/dE_{\rm rel}}
\newcommand{\vecrone}{\vec{r_1}}
\newcommand{\vecrtwo}{\vec{r_2}}
\newcommand{\rcn}{\vec{r}_{cn}}
\newcommand{\rcnn}{\vec{r}_{c-nn}}
\newcommand{\rnn}{\vec{r}_{nn}}
\newcommand{\boldr}{\mbox{\boldmath$r$}}
\newcommand{\boldk}{\mbox{\boldmath$k$}}
\newcommand{\boldK}{\mbox{\boldmath$K$}}

\section{Dineutron for two-neutron halo nuclei}\label{sec:di}

The neutron halo is a peculiar nuclear-structure feature, in which one or two valence neutrons extend far outside the core's mean-field potential, forming a low-density neutron cloud. This feature occurs for
weakly bound neutron(s), whose one- or two-neutron separation energies, $\sn$ (or $\snn$), are less than about 1 MeV with low orbital angular momentum, $\ell=0,1$. The known ground-state neutron-halo nuclei 
%\red{({\bf ALTERNATIVE:} The known nuclei having the neutron-halo character in the ground state)} 
thus lie along the neutron drip line on the nuclear chart.
Two-neutron halo nuclei are composed of the core and the two halo neutrons: for example, $^{11}$Li, 
the first halo nucleus observed~\cite{TANI85}, has a three-body structure composed of the $^{9}$Li core and the two halo neutrons. 
More comprehensive reviews on halo nuclei are available in Refs.~\cite{TANI13, TANI20}.

\begin{figure}[h]
 \begin{center}
   \includegraphics[width=8.cm]{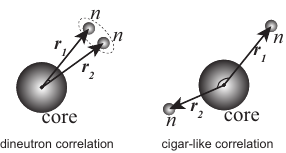}
   \caption{Two types of $nn$ correlations, dineutron~(left), and cigar-like correlations~(right), which are predicted for typical halo nuclei, such as $^{11}$Li and $^{6}$He.
   The figure is reproduced from Ref.~\cite{NAKA23} with permission from Springer Nature}
   \label{fig:correlation}
   \end{center}
\end{figure}

Any known two-neutron halo nuclei, $^{6}$He, $^{11}$Li, $^{14}$Be, $^{17,19}$B, $^{22}$C, and $^{29}$F, have a three-body Borromean character, where the two-body constituents ($n$-$n$, and $n$-core) are unbound, while the whole three-body system ($n$-$n$-core) is bound. In this sense, the $nn$ correlation in two-neutron halo nuclei plays a significant role in their stability.   
It is also noted that the Borromean two-neutron halo system 
is closely related
to the more universal Efimov physics~\cite{NAID17,NAID23}.
 
Possible geometries of a two-neutron halo nucleus are shown in Fig.~\ref{fig:correlation}: dineutron- and cigar-like configurations. 
This dual structure was predicted by calculations of the three-body model for the $p$-shell two-neutron halo nuclei~\cite{ZHUK93,HASA07_2}. The dineutron correlation represents a spatially compact two-neutron system, where the opening angle between $\boldr_1$ and $\boldr_2$ is significantly less than $90^\circ$, while the cigar-like correlation is the state for the two neutrons on the opposite side with respect to the core (opening angle $\sim 180^\circ$). The existence of dineutron correlation in two-neutron halo nuclei has been studied experimentally. The main experimental methods are Coulomb breakup, charge-radius measurements, and quasi-free proton scattering via the $(p,pn)$ reaction, as shown below.
Theoretical studies on dineutron correlation are also shown.

\subsection{Experimental studies of dineutron in two-neutron halo nuclei}
\label{sec:diexp}
\subsubsection{Coulomb breakup of two-neutron halo nuclei and dineutron correlation}\label{sec:coul}
Coulomb breakup is a process for a fast projectile (rare-isotope beam) to be excited by the absorption of a virtual photon
when it passes by a high-$Z$ target, such as Pb and Au, and to undergo breakup as the excitation energy exceeds the breakup threshold~\cite{BERT88,AUMA13,NAKA23}.
Coulomb breakup is the most suitable tool for investigating the electric dipole response ($\eone$) of halo nuclei, since halo nuclei have a feature called soft $\eone$ excitation,
which is the low-energy $\eone$ response arising from the displacement of the center of the halo neutrons from the center of the core in its intrinsic frame.
 Furthermore, the 
halo wave function outside the core, coupled with the $\eone$ operator, and the low-energy continuum strongly overlap,
resulting in strong $\eone$ strength at low relative energies.
Such a mechanism for one-neutron halo nuclei is shown in the experiments for $^{11}$Be~\cite{NAKA94,FUKU04,PALI03,AUMA13,NAKA23},
and supported by theoretical works~\cite{OTSU94,NAGA05}.
As such, the soft $\eone$ excitation is generally considered to be predominantly non-resonant.

The Coulomb breakup process can be well approximated by the equivalent photon method. There, the Coulomb breakup (CB) energy differential cross section, $\dsderelone$, is directly related to the $\eone$ reduced transition probability, $\dbderelone$, 
 \begin{equation}
\dsderel=\frac{16 \pi^3}{9\hbar c}\neone\dbderel,
\label{eq:cb}
\end{equation}
where $\erel$ is the relative energy of the breakup particles, $\ex$ is the excitation energy,
and $\neone$ is the $\eone$ virtual photon number for a given photon energy ($=\ex$)~\cite{BERT88,AUMA13,NAKA23}. 
For a two-neutron halo nucleus with the two-neutron separation energy, $\snn$, 
$\erel$ is the three-body relative energy for the core $+n+n$ and is related to $\ex$ as $\erel=\ex-\snn$.
Eq.(\ref{eq:cb}) illustrates that $\dsderelone$ is amplified by $\neone$,
which decreases rapidly as $\ex$ increases. 
The soft $\eone$ excitation is thus probed more efficiently as $\neone$ is predominant for lower excitation energies. 

\begin{figure}[ht]
\centering
\includegraphics[width=80.mm]{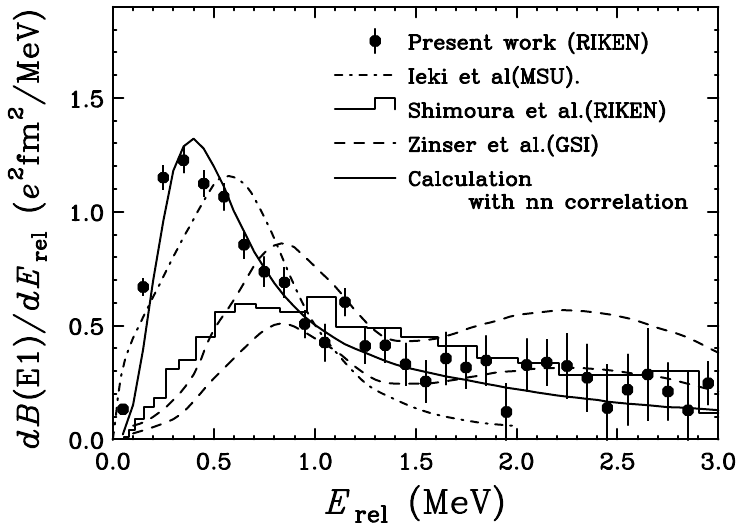}
\caption{$\beone$ distribution of $^{11}$Li as a function of the $^9$Li+$n$+$n$ relative energy, $\erel$. A strong electric dipole strength is found to be concentrated below $\erel< 1$ MeV, typical of the soft $\eone$ excitation. The solid circles are the experimental points, while the dot-dashed curve, solid histogram, and dashed curves are from the previous measurements, which may have been affected by $\erel$ dependent efficiencies, whose detail is shown in Ref.~\cite{NAKA06}. The solid curve is the prediction by the three-body model~\cite{ESBE92}. The figure is reproduced from Ref.~\cite{NAKA06}}\label{fig:beone11Li}
\end{figure}

Figure~\ref{fig:beone11Li} shows the experimental $\beone$ distribution for $^{11}$Li obtained by using its Coulomb breakup with a Pb target at 70 MeV/nucleon~\cite{NAKA06}.
The spectrum shows a strong peak at low energies, with $\erel\sim 0.3$~MeV. Note that $\snn$ of $^{11}$Li is only 0.3693(6) MeV~\cite{SMIT08,AME2020}, so the excitation energy is also very small.
This is a typical soft $\eone$ excitation spectrum in neutron halo nuclei.
Compared is the prediction of the three-body model by Esbensen~\cite{ESBE92}, which reproduced the data very well.

\begin{figure}[ht]
\centering
\includegraphics[width=40.mm]{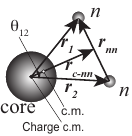}
\caption{The three-body geometry of a two-neutron halo nucleus. The figure is reproduced from Ref.~\cite{NAKA23} with permission from Springer Nature}\label{fig:halo_coord}
\end{figure}

The integrated $\beone$ has a direct relation to the geometry of a two-neutron halo nucleus.
The $\eone$ non-energy weighted cluster sum rule reads
 \begin{eqnarray}
\beone = \int_{-\infty}^\infty{\frac{d\beone}{d\ex}}d\ex
          &=&\frac{3}{4\pi} \left(\frac{Ze}{A} \right)^2
            \langle (\vecrone+\vecrtwo)^2 \rangle, \nonumber \\
            &=& \frac{3}{4\pi} \left(\frac{Ze}{A} \right)^2
            \langle r_1^2+r_2^2+ 2\vecrone\cdot \vecrtwo \rangle, \nonumber \\
            &=& \frac{3}{\pi}\left(\frac{Ze}{A} \right)^2 \langle r_{c-nn}^2 \rangle,
            \label{eq:eonesum2}
\end{eqnarray}
where $\vecrone$ and $\vecrtwo$ are the position vectors of the respective two halo neutrons relative to the center-of-mass (c.m.) of the core, as shown in  Fig.~\ref{fig:halo_coord}. $\vec{r}_{c-nn}\equiv(\vec{r}_1+\vec{r}_2)/2$ is the distance between the c.m. of the two valence neutrons and that of the core nucleus. 
Hence, the $\beone$ cluster sum rule for two neutron halo
nuclei involves geometric information, such as $\sqrt{\langle r_{c-nn}^2 }\rangle$, and
the correlation $\langle\vecrone\cdot\vecrtwo\rangle$. In Ref.~\cite{NAKA06}
the mean opening angle $\langle \theta_{12} \rangle$  (Fig.\ref{fig:halo_coord})
was evaluated to be $48^{+14}_{-18}$ degrees and 
$\sqrt{\langle r_{c-nn}^2\rangle}=5.01(32)$~fm. 
We should note that $\sqrt{\langle r_{c-nn}^2 }\rangle$
is 
extracted directly from the integrated $\beone$, while for 
$\langle \theta_{12}\rangle$ additional input is needed. In Ref.\cite{NAKA06}, the theoretical model for the non-correlated $\beone$~\cite{ESBE92} value was used for this evaluation.

Refs.~\cite{HASA07,BERT07,ESBE07} introduced a simple relation (shown below in Eq.~(\ref{eq:rnn})) that the information on the rms matter radii of the two-neutron halo nucleus and the core, combined with $\langle r^2_{c-nn} \rangle$, will solve this ambiguity about the three-body geometry. 
When we look into the geometry shown in Fig.~\ref{fig:halo_coord}, the mean square matter radius of the halo nucleus is related to the mean square radii of the core and the halo measured from the c.m. of the entire system,
 as,
\begin{equation}
A \langle r_m^2 \rangle = A_c \langle r_m^2 \rangle_{c(cm)}
+ 2 \langle r_m^2 \rangle_{h(cm)},    
\label{eq:matter}
\end{equation}
where $\langle r_m^2 \rangle_{c(cm)}$ and $\langle r_m^2 \rangle_{h(cm)}$ are, respectively, the mean-square matter radius of
the core and the halo with respect to the c.m. of the entire nucleus (c.m. in Fig.\ref{fig:halo_coord}). 
$A$ and $A_c$ are the mass numbers of the halo nucleus and the core ($A=A_c+2$), respectively.
Here,
\begin{equation}
    \langle r_m^2 \rangle_{c(cm)}= \langle r_m^2 \rangle_{c} + 
    \left(\frac{2}{A}\right)^2\langle r^2_{c-nn} \rangle,
    \label{eq:matter_core}
\end{equation}
where $\langle r_m^2 \rangle_{c}$ represents the mean square radius of
the core with respect to the c.m of the core (equivalently, charge c.m.). 
On the other hand, for the halo,
\begin{equation}
    \langle r_m^2 \rangle_{h(cm)}= \langle \left(\frac{r_{nn}}{2}\right)^2 \rangle + 
    \left(\frac{A_c}{A}\right)^2\langle r^2_{c-nn}\rangle,
    \label{eq:matter_halo}
\end{equation}
where $r_{nn}$ is the distance between the two halo neutrons shown in Fig.~\ref{fig:halo_coord}. 
Inserting Eqs.~(\ref{eq:matter_core}) and (\ref{eq:matter_halo}) into
Eq.~(\ref{eq:matter}), we obtain the following useful relation,
\begin{equation}
\langle r_m^2 \rangle=\frac{A_c}{A}\langle r_m^2 \rangle_c
+ \frac{2A_c}{A^2}\langle r^2_{c-nn}\rangle + \frac{1}{2A}\langle r^2_{nn} \rangle.
\label{eq:rnn}
\end{equation}
This equation demonstrates that the rms matter radii of the halo nucleus and its core, combined with the rms $r_{c-nn}$ value, provide the rms $nn$ distance. This means that the three-body geometry of the two-neutron halo nuclei shown in Fig.\ref{fig:halo_coord} can be determined. 
Using the relations,
\begin{eqnarray}
\vec{r_1} &=& \vec{r_{c-nn}} + \frac{\vec{r_{nn}}}{2},\\
\vec{r_2} &=& \vec{r_{c-nn}} - \frac{\vec{r_{nn}}}{2},
\end{eqnarray}
together with the assumption that $\langle r_1 \rangle = \langle r_2 \rangle\equiv\langle r\rangle$, one obtains $\langle r^2\rangle$ as a function of $\langle r_{c-nn}^2\rangle$ and $\langle r_{nn}^2\rangle$. This assumption is a good approximation in the three-body model, where $\vec{r}_{c-nn}$ and $\vec{r}_{nn}$ are, on average, perpendicular to each other.
The mean opening angle between the two neutrons, $\langle \theta_{12} \rangle$, a key observable for the dineutron correlation, is then determined using
\begin{equation}
\langle\cos\theta_{12}\rangle =1- \frac{\langle r_{nn}^2\rangle}{2\langle r^2 \rangle}.
\label{eq:cosonetwo}
\end{equation}

%The mean opening angle of the two neutrons, $\langle \theta_{12} \rangle$, a key observable for the dineutron correlation, is thus determined.

Using Eqs.~(\ref{eq:rnn}) and (\ref{eq:cosonetwo}), Ref.~\cite{HASA07} extracted 
$\sqrt{\langle{r^2_{nn}}\rangle}=5.9(1.2)$~fm, and $\langle\theta_{12}\rangle=56.2^{+17.8}_{-13.1}$ degrees, respectively.
In the same reference, they also apply
the $\sqrt{\langle r^2_{nn} \rangle}$ value, derived from the Hanbury Brown and Twiss~(HBT) analysis of the 
$^{11}$Li three-body breakup~\cite{MARQ00}, for the estimation of the opening angle. However, we note that the HBT method may be affected by final-state interactions (FSI), and the correlation may not correspond to the ground state of $^{11}$Li. Ref.~\cite{BERT07} also estimated the opening angle based on such an analysis.

The mean opening angle $\langle \theta_{12}\rangle <90^\circ$ indicates that there is a dineutron correlation in $^{11}$Li. A detailed theoretical interpretation is presented in Sec.~\ref{sec:ditheo}.

We note that we require some assumptions 
in extracting the correlation angle from the integrated $\beone$. 
%Firstly,  we assume that $\langle r_1 \rangle = \langle r_2 \rangle$, so that $\vec{r}_{c-nn}$ and $\vec{r}_{nn}$
%are perpendicular to each other. 
%This is a good approximation based on the three-body model calculation.
In addition to the assumption that $\langle r_1 \rangle = \langle r_2 \rangle$, we assume that the $\beone$ strength we adopt
is exhausted by the non-energy weighted cluster sum rule. This also assumes that the two-neutron halo nucleus is well described by the three-body system: core and the two neutrons.
In the case of $^{11}$Li~\cite{NAKA06}, we also needed to 
extrapolate the $\beone$ distribution to higher energies, as the experiment extracted $\beone$ up to $\erel=$3 MeV, and the distribution above 3 MeV assumes the one predicted by Esbensen~\cite{ESBE92}.

\begin{figure}[ht]
\centering
\includegraphics[width=70.mm]{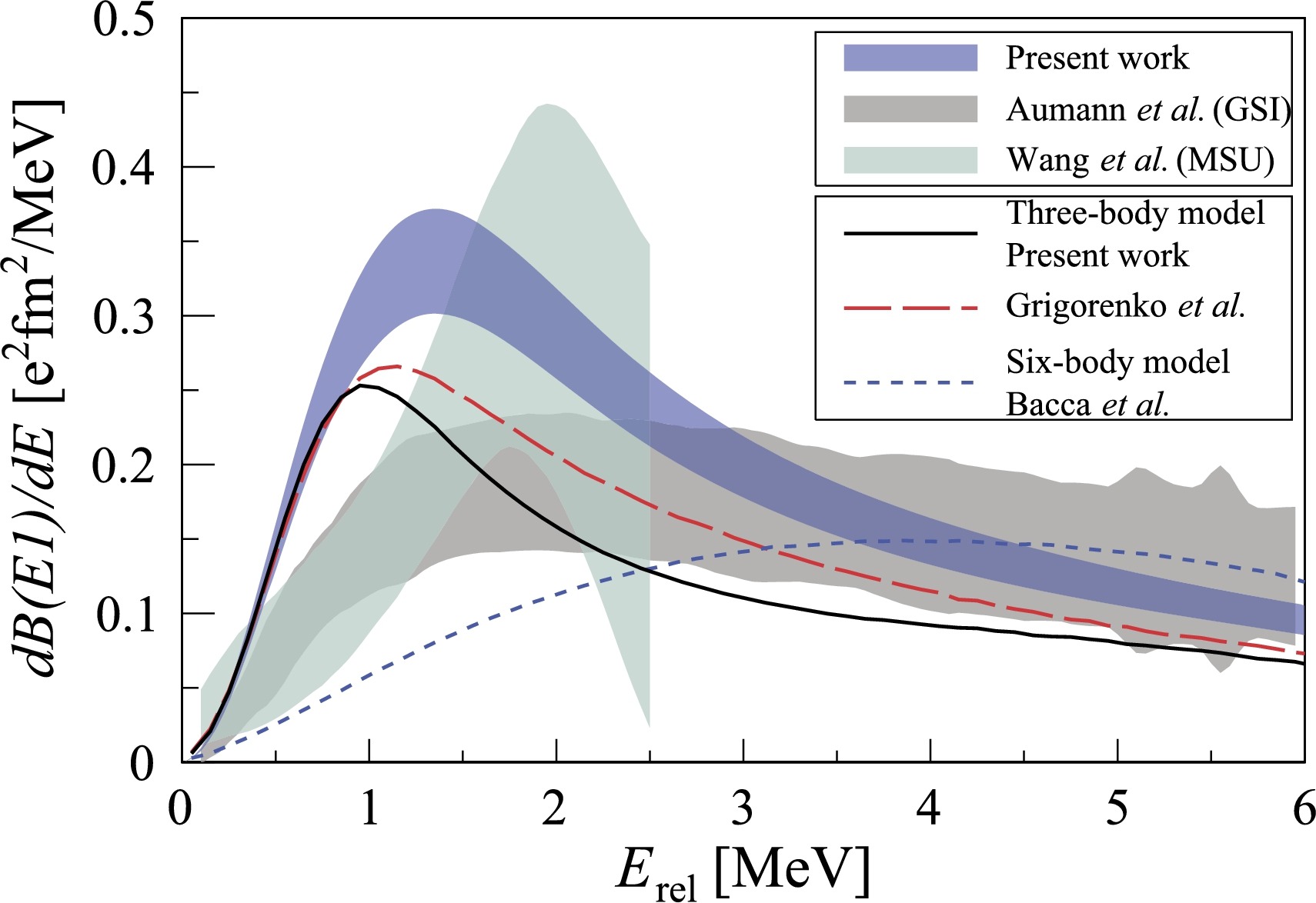}
\caption{$\beone$ distribution of $^{6}$He as a function of the $^4$He+$n$+$n$ relative energy, $\erel$. The bluish-violet shaded area shows the CDCC results (parameters optimized to reproduce the experimental $\dsderel$ of $^6$He+Pb at 70 MeV/nucleon). For details of the other curves, see Ref.~\cite{SUN21}. The figure is reproduced from Ref.~\cite{SUN21} under the Creative Commons CC-BY license}\label{fig:beone6he}
\end{figure}

For $^6$He, Fig.~\ref{fig:beone6he} shows the result of the Coulomb breakup experiment at 70~MeV/nucleon at RIKEN~\cite{SUN21}.
Here, the $\beone$ distribution that best reproduces the experimental energy-differential cross sections is extracted from the continuum discretized coupled channels (CDCC) calculation, as shown by the bluish-violet band in the figure.
The integrated $\beone$ for $\erel\leq~20$~MeV for this estimation amounts to 1.6(2)~$e^2$fm$^2$, which corresponds to the rms core-$nn$ distance being $\sqrt{\langle r^2_{c-nn}\rangle}=3.9(2)$~fm. Using Eq.~(\ref{eq:rnn}),
with rms matter radii for $^6$He (2.49(4)~fm) and for $^4$He (1.463(6)~fm), $\sqrt{\langle r^2_{nn}\rangle}$=4.1(7)~fm, and
$\langle\theta_{12}\rangle$=56$^{+9}_{-10}$ degrees were extracted.
As this angle is narrower than 90$^\circ$, this provides evidence of dineutron correlation in $^6$He.
It is interesting to note that the mean opening angle is nearly the same between $^{11}$Li and $^6$He, where the two neutron configurations differ significantly. For $^{11}$Li, the valence neutrons are a mixture of dominant $(1p)^2$ and $(2s)^2$, while for $^6$He, they are predominantly $(1p)^2$. 

For the $^{19}$B Coulomb breakup at 220 MeV/nucleon, 
$\beone$ is extracted as shown in Ref.\cite{COOK20}.
There, the integrated $\beone$ value, 1.64$\pm$0.06(stat)$\pm$0.12(sys) $e^2$fm$^2$, and the resultant rms core-nn distance, $\sqrt{\langle r^2_{c-nn}\rangle}$=5.75$\pm$0.11(stat)$\pm$0.21(sys)~fm were extracted. However, due to the large uncertainties of the matter radii of $^{19}$B and $^{17}$B~\cite{SUZU99}, the opening angle was not extracted. In the future, more accurate extraction of such radii is needed.
On the other hand, as demonstrated in Ref.\cite{COOK20}, the dineutron correlation in $^{19}$B is suggested by the three-body model calculation to best reproduce the $\beone$ distribution.

\subsubsection{Charge radii  of two-neutron halo nuclei and dineutron correlation}\label{sec:chradii}

The charge radius can also provide information on the dineutron correlation. This is because the rms charge radius
can be correlated not only with the core radius but also with the shift of the core c.m. from the c.m. of the entire nucleus
due to the spatial $nn$ correlations. 

In general, the mean square charge radius of a nucleus, $\langle r_{ch}^2 \rangle$,  is directly related to that of its proton density distribution as
\begin{equation}
\langle r_{ch}^2 \rangle = \langle r^2_{p} \rangle +
\langle R^2_{p} \rangle + \frac{N}{Z}\langle R^2_{n} \rangle
+ \frac{3\hbar^2}{4m_p^2 c^2},
\label{eq:charger}
\end{equation}
where $\langle r_p^2 \rangle$ is the mean-square radius of the point proton distribution, $\langle R_p^2 \rangle$ ( $\langle R_n^2 \rangle$ ) is the mean-square proton (neutron) charge radius, and the last term is the Darwin-Foldy term (0.033 fm$^2$)\cite{TANI13,FRIA97}.  The current particle data group evaluates the proton rms radius, $\sqrt{\langle R_p^2 \rangle}=0.8409(4)$~fm, and the neutron {\it mean-square} radius,  $\langle R_n^2 \rangle=-0.1155(17)$~fm$^2$~\cite{PDG2024}.

For a two-neutron halo nucleus, the rms radius of the proton distribution corresponds to that of the core nucleus 
when the c.m. of the two halo neutrons matches that of the core.
However, when a dineutron correlation occurs, and this condition is not met, the rms radius of the proton distribution increases. 
Therefore, information on the dineutron correlation can be extracted from the rms proton distribution.

Quantitatively, this can be understood in a similar way for the Coulomb breakup experiments ($\beone$) shown in Eq.~(\ref{eq:matter_core}). 
The mean-square radius of the proton distribution, 
$\langle r_{p}^2 \rangle$, is expressed as
\begin{equation}
\langle r_{p}^2 \rangle \equiv \langle r_{p}^2 \rangle_{c(cm)}= \langle r_p^2 \rangle_{c} + 
    \left(\frac{2}{A}\right)^2\langle r^2_{c-nn} \rangle,
\end{equation}
where $\langle r_{p}^2 \rangle_{c(cm)}$ expresses 
the mean-square proton-distribution radius
measured in the c.m. of the entire system, and  $\langle r_{p}^2 \rangle_c$
is the mean-square proton-distribution radius of the core.

Consequently, the charge radius in Eq.~(\ref{eq:charger}) can be related to the rms distance between the core and the c.m. of the two halo neutrons, $\sqrt{\langle r^2_{c-nn}\rangle}$, in 
Fig.~\ref{fig:halo_coord}, as in the case of low lying $\beone$ strength (soft $\eone$ excitation).
Once we obtain $\sqrt{\langle r^2_{c-nn}\rangle}$, 
combining this with the matter radii
of the core nucleus and the whole three-body system,
one can obtain $\sqrt{\langle r^2_{nn}\rangle}$ using
Eq.~(\ref{eq:rnn}). The mean opening angle, $\langle\theta_{12}\rangle$, can also be extracted.

Experimentally, the charge radii have been extracted 
for stable nuclei via electron scattering.
However, for unstable nuclei, such as neutron-halo nuclei, electron-scattering experiments are not feasible with current technology.
Instead, isotope shift measurements provide an alternative, powerful tool for extracting rms charge radii of nuclei. 
In the future, SCRIT technology, which was demonstrated to work for unstable nuclei at RIKEN\cite{TSUK17,TSUK23}, 
may be applicable to extract charge radii of halo nuclei by electron scattering. 
%But this is still a decade away.

The isotope shift is a minute difference 
in atomic transition energies between isotopes. 
The shift is attributed to the mass shift and the field shift:
\begin{equation}
\delta \nu = \delta \nu_{MS} + \delta \nu_{FS},
\end{equation}
where $\delta \nu$ is the difference in the
frequency of the same atomic transitions 
between isotopes: the isotope shift to be measured experimentally.
$\delta \nu_{MS}$ represents the mass shift, which is attributed to the atomic reduced mass,
while $\delta \nu_{FS}$ represents the field shift, which is attributed to the change of the electric charge in the nucleus. Since $\delta\nu_{MS}$ can theoretically be determined precisely enough,  $\delta\nu_{FS}$ can be extracted from $\delta \nu$, and is related to the rms charge radius:
\begin{equation}
\delta \nu_{FS}\propto Z\times\Delta|\phi(0)|^2\times\delta \langle r^2_{ch}\rangle,
\end{equation}
where $\Delta|\phi(0)|^2$ represents the difference in the electron wave function in the nucleus and $\delta\langle r^2_{ch}\rangle$ is the difference in the root mean square charge radii of the nucleus. As such, the rms charge radius of exotic nuclei can be extracted.

The isotope shifts for lithium isotopes were measured using high-resolution laser spectroscopy for two-photon Doppler-free excitation~\cite{SANC06, NORT11}. For neutron-rich lithium isotopes, such as $^{11}$Li and $^9$Li, their isotope shifts relative to $^{7}$Li were successfully measured at TRIUMF~\cite{SANC06}.
The extracted charge radius of $^{11}$Li, $\sqrt{\langle r^2_{ch}\rangle}$($^{11}$Li)=2.467(37)~fm, was significantly larger than that of $^9$Li, 
$\sqrt{\langle r^2_{ch}\rangle}$ ($^{9}$Li)=2.217(35) fm.
This enhancement of the charge radius in $^{11}$ Li can be attributed to the dineutron correlation in the halo.

Ref.~\cite{BERT07} evaluated $\sqrt{\langle r^2_{c-nn} \rangle} =5.97(22)$~fm for $^{11}$Li from the charge radius measurement~\cite{SANC06}, which is greater than 5.01(32)~fm evaluated from the Coulomb breakup experiment~\cite{NAKA06}.
This may show an even stronger dineutron correlation in $^{11}$Li.
Hagino and Sagawa re-evaluated the latter value to be 5.15(33)~fm using a more realistic extrapolation method to higher excitation energies in the $\beone$ distribution. However, a discrepancy of about 1.5$\sigma$ remains. 
Ref.~\cite{ESBE07} discusses this discrepancy and suggests that this might be due to the core polarization effect. We should note that the discussion here is based on the three-body model of the two-neutron halo system, in which the core nucleus is identical to its bare state. The discrepancy may show a hint of a change in the core state inside the halo nucleus.
We also note that the extrapolation of the $\beone$ distribution by the theories may have some uncertainties, which should be investigated further. 
The experimental $\beone$ distribution for higher excitation energy would also be needed.

For helium isotopes, the isotope shift for $^6$He relative to $^4$He
was first measured at Argonne National Laboratory~\cite{WANG04}, where the $^6$He atoms were confined and cooled in a magneto-optical trap (MOT), and high-precision laser spectroscopy was performed. In the subsequent experiment at GANIL, the isotope shifts of both $^6$He and $^8$He were successfully measured~\cite{MUEL07}. We should note that to evaluate the charge radii, a precise theoretical evaluation of the mass shift, $\delta_{MS}$, is essential as $\delta_{MS}\gg\delta_{FS}$ 
for light nuclei. Recent advances in the theory of the atomic structure of helium played a crucial role in the extraction of 
$\delta_{MS}$ with ultra-high precision~\cite{WANG04}.

The extracted rms charge radii of $^{6}$He and $^8$He were  2.068(11)~fm and 1.929(26)~fm, compared to that of $^4$He, 1.676(8)~fm ~\cite{SICK82}, respectively. It should be noted that the smaller radius of $^{8}$He compared to $^6$He can be attributed to the fact that $^6$He has a dineutron halo structure that enhances the charge radius, while for $^8$He, four valence neutrons tend to be distributed in a more spherically symmetric way\cite{MUEL07}, with a possible two-dineutron structure~\cite{KANA07,HAGI08,ITAG08,KOBA13b,YAMA23,NAKAG25}.

Ref.~\cite{BERT07} evaluates the rms $r_{c-nn}$ of $^6$He to be 3.71(07)~fm using the first charge radius measurement~\cite{WANG04}(2.054(14)~fm). Interestingly, this value is consistent with the value 3.9(2)~fm from the Coulomb breakup~\cite{SUN21}. 
This agreement may support the hypothesis that the discrepancy in the rms $r_{c-nn}$ values observed for $^{11}$Li could be due to core polarization or a change in the core size. 
For $^6$He, on the other hand, the core polarization of $^{4}$He should be negligible because of the inertness of $^{4}$He.
Tanihata et al. pointed out that the discrepancy of rms $r_{c-nn}$ between the derivations from the charge radius and the Coulomb breakup is also found for the one-neutron halo nucleus $^{11}$Be~\cite{TANI13}.
They examined the possibility of changing the core size and found about a 2\% increase. 
Further experimental and theoretical investigations on this issue should be extended to other one-neutron and two-neutron halo nuclei in the future.

\subsubsection{Quasi-free scattering of two-neutron halo nuclei and dineutron correlation}\label{sec:quasi}

We expect that the dineutron correlation appears in a specific low-density region of a nucleus: Matsuo predicted that the dineutron correlation becomes significant for the nuclear density of $10^{-4}<\rho/\rho_0<0.5$~\cite{MATS06}, as discussed in Sec.~\ref{sec:intro}.
However, Coulomb breakup and charge radii experiments can
provide only a mean opening angle, $\langle\theta_{12}\rangle$.
We therefore need experimental methods to study density-dependent $nn$
correlation, and proton quasi-free scattering, $(p,pn)$, of a two-neutron halo nucleus is one solution. This reaction can provide not only the mean opening angle of the two valence neutrons but also their nuclear-density dependence, which is discussed here. 

\begin{figure}[ht]
\centering
\includegraphics[width=\linewidth]{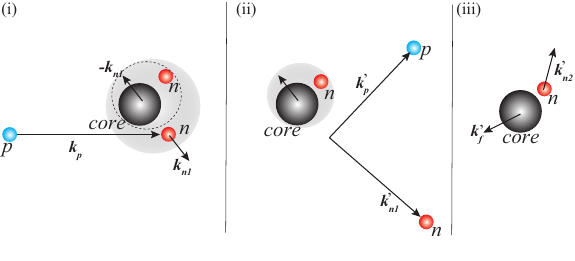}
\caption{Kinematics of the $(p,pn)$ reaction in normal kinematics: (i) before the collision, (ii) immediately after the collision, and (iii) decay of the residual}\label{fig:quasifree}
\end{figure}

\begin{figure}[ht]
\centering
\includegraphics[width=100.mm]{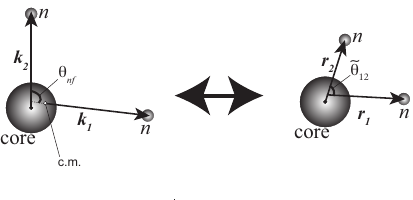}
\caption{Correlation angles for the two valence neutrons: (left) $\theta_{nf}$ in the momentum space, defined in the Jacobi $\mathbf{Y}$ coordinate system; (right) $\tilde{\theta}_{12}$ in the  Jacobi $\mathbf{Y}$ system.}\label{fig:quasifreeco}
\end{figure}

Figure~\ref{fig:quasifree} schematically shows the kinematics of the proton quasifree scattering with a two-neutron halo nucleus in normal kinematics: the halo nucleus at rest is hit by a proton with incident momentum $\boldk_p$.
Before collision (i), one valence neutron ($n_1$) in the halo has a momentum $\boldk_{n1}$. Here, this neutron is a participant, and the other part of the nucleus (core+$n_2$) is a spectator: the neutron in the halo ($n_1$) is elastically scattered by the incident proton.
After collision (ii), the proton and neutron are emitted with an opening angle of about $90^{\circ}$, when $k_{n1}$ is sufficiently smaller than $k_p$. In the experiment, we measure the energies and directions of the proton and neutron, $\boldk'_p$ and $\boldk'_{n1}$, respectively.
The decay of the residue into the core fragment and the neutron follows (iii).

The momentum conservation provides the momentum of $n_1$
inside the nucleus before the collision:
\begin{equation}
\boldk_1:=\boldk_{n1}=\boldk'_{n1}+\boldk'_p-\boldk_p.
\end{equation}
The momenta of the core $\boldk'_f$ 
and neutron $\boldk'_{n2}$ provide an approximate momentum 
of $n_2$ in the rest frame of the core+$n_2$ as 
\begin{equation}
\boldk_2 \approx\boldK'=\boldk'_{n2}-\boldk'_{f}.     
\end{equation}
The opening angle between $\boldk_1$ and $\boldK'$ provides
an estimate of the correlation angle between the two valence neutrons in the momentum space, as depicted in Fig.\ref{fig:quasifreeco}~(left).
Here, the Jacobi $\mathbf{Y}$ system is adopted.
The opening angle $\theta_{nf}$ can be extracted by using
\begin{equation}
\cos{\theta_{nf}}\approx\frac{\boldk_1\cdot\boldK'}{|\boldk_1||\boldK'|}.
\end{equation}
As shown in Fig.\ref{fig:quasifreeco}, $\langle \theta_{nf} \rangle$ is momentum conjugate to $\langle \tilde{\theta}_{12}\rangle$, which is the opening angle in the Jacobi $\mathbf{Y}$ coordinate system. Consequently, 
\begin{equation}
\langle \theta_{nf} \rangle\sim 
180^{\circ}-\langle \tilde{\theta}_{12}\rangle.
\end{equation}
Note that $\theta_{12}\approx\tilde{\theta}_{12}$ for a heavy core, where $\theta_{12}$ is the correlation angle defined relative to the core center as shown in Fig.~5.

In the $^{11}$Li experiment, we use inverse kinematics because $^{11}$Li is provided as a beam. The pioneering experiment that attempted to extract $\langle\theta_{nf}\rangle$ used a carbon target instead of a proton target~\cite{SIMO99}. This experiment applied the fact that $\boldk_{n1}$ can be extracted from the sum of the momenta of the core and the second neutron. The mean opening angle, 103.4(2.1) degrees, was thus extracted. The angle greater than 90 degrees suggested the presence of dineutron correlation. However, the density dependence was not discussed.

Recently, a kinematically complete measurement of the reaction $^{11}$Li($p,pn$) in inverse kinematics at 246 MeV/nucleon~\cite{KUBO20} was performed at the SAMURAI facility~\cite{KOBA13} at RIBF, RIKEN.
In the experiment, the recoiled proton, the knocked-out neutron at $\sim$45 degrees, $^9$Li and the second neutron at the forward angles were measured in coincidence. The proton target has the advantage that it can probe the entire volume of $^{11}$Li, while the carbon target tends to probe the surface of the nucleus.   

A new target system, MINOS, consisting of a 15~cm-thick liquid-hydrogen target combined with the surrounding time-projection chamber, played an essential role in enhancing the event rate while maintaining good energy resolution~\cite{SANT18, OBER11}. 
Since the vertex point in the target can be detected event-by-event, the energy-loss correction can be applied to maintain good energy resolution. This device has made the experiments at SAMURAI very productive, where a series of proton-knockout reactions have been performed, including the observation of $^{28}$O~\cite{KOND23} as later described. 

\begin{figure}[ht]
\centering
\includegraphics[width=90.mm]{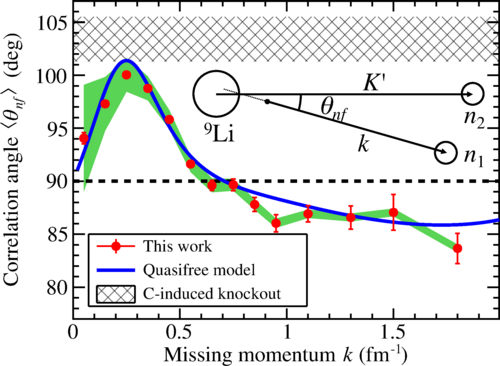}
\caption{Mean $nn$ correlation angles in momentum space (Jacobi $\bold{Y}$ system),$\langle \theta_{nf}\rangle$, as a function of missing momentum $k$.
$\langle \theta_{nf}\rangle$ peaks at $k\sim 0.25$ fm$^{-1}$ and is larger than 90$^\circ$, indicating that the dineutron correlation is strong on the surface of the 
core.
The figure is reproduced from Ref.~\cite{KUBO20}
}\label{fig:kubota}
\end{figure}

Figure~\ref{fig:kubota} shows the mean correlation angle, $\langle \theta_{nf}\rangle$, as a function of 
the missing momentum $k$, which represents the opening angle between
$\boldk(=\boldk_1)$ and $\boldK'(\approx \boldk_2)$ as
shown in this figure (see also Fig.~\ref{fig:quasifreeco}~(left)). The peak appears at $k=0.25$~fm$^{-1}$ with $\langle \theta_{nf}\rangle\approx 100^\circ$, larger than 
the non-correlated angle of $90^\circ$, exhibiting the dineutron correlation. On the other hand, the region with $k > 0.7$~fm$^{-1}$ shows smaller angles, implying that no dineutron correlation appears at such high $k$.
Since $k$ represents the momentum of a halo neutron in the ground state of $^{11}$Li, it can be associated with the corresponding density region. 
That is, the distribution shows that the dineutron is formed in a specific density region, 
$10^{-3}\lesssim \rho/\rho_0 \lesssim 10^{-2}$, as estimated using the quasi-free model by Kikuchi et al.~\cite{KIKU16}.
A simple Fermi-gas model ($\rho=k^3/(3\pi^2)$) also yields a density range of the same order.
This density region for the dineutron is located near the surface of the $^9$Li core.  Note that the dineutron correlation is weaker not only inside the core (higher $k$, higher density), but also far away from the core ($k\sim$~0, very low density). The latter can be understood by the fact that the two neutrons lying farther from the core interact more weakly. 
It is interesting that this finding is qualitatively consistent with Matsuo's theoretical prediction using the mean field model for the nuclear matter, shown in Fig.\ref{fig:densedep} : the dineutron is predicted to occur for the limited density region with $10^{-4}\lesssim \rho/\rho_0 \lesssim 0.5$~\cite{MATS06}.

\begin{figure}[htb]
 \begin{center}
   \includegraphics[width=9cm]{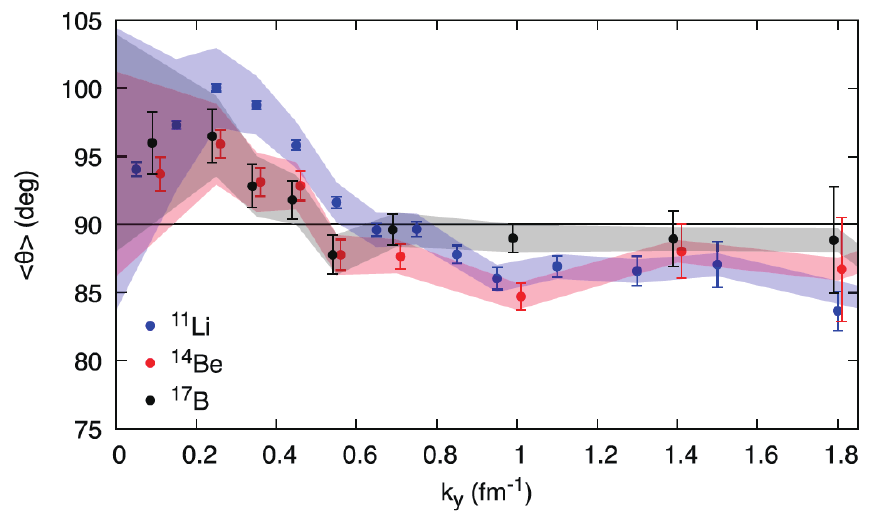}
   \caption{Mean correlation angles, $\langle \theta \rangle$, in momentum space obtained by the $^{11}$Li,$^{14}$Be,$^{17}$B($p,pn$) reactions as a function of missing momentum of the neutron inside a nucleus, $k_y$.  $\langle \theta \rangle$ and $k_y$ correspond to $\langle \theta_{nf} \rangle$ and $k$ used in the discussion of Fig.~\ref{fig:kubota}, respectively. The peaks for the dineutron correlation appear around 0.25~fm$^{-1}$ in all these nuclei.  The figure is reproduced from Ref.~\cite{CORS23} under the Creative Commons CC-BY license
   }
   \label{fig:corsi}
   \end{center}
\end{figure}

Figure~\ref{fig:corsi} shows the correlation angles in momentum space
obtained for Borromean nuclei, $^{14}$Be and $^{17}$B, in addition to $^{11}$Li~\cite{CORS23}. The data for $^{11}$Li is from Ref.\cite{KUBO20}. 
Here, $\langle \theta \rangle$ and $k_y$
correspond to $\langle \theta_{nf} \rangle$ and $k$ in Fig.~\ref{fig:kubota},
respectively.
Interestingly, the correlation angles for $^{14}$Be and $^{17}$B 
both peak at $k_y\sim 0.25$~fm$^{-1}$ as in the case of $^{11}$Li. 
We should note that the configurations of the two valence neutrons in $^{11}$Li, $^{14}$Be, and $^{17}$B vary significantly. 
For $^{11}$Li, the configuration ratio is ($1s_{1/2})^2 : (0p_{1/2}) : (0d_{5/2})^2 = 
35(4)\%: 59(1) \% : 6(4) \%$~\cite{KUBO20}, while that for $^{14}$Be is 19\% : 61\% : 20\%, respectively, which contains about 20\% 2$_1^+$ core-excited component~\cite{CORS21}. For $^{17}$B, Ref.~\cite{YANG21} shows that $(1s_{1/2})^2$ configuration occupies only 9(2)\%, while the remaining is dominated by $(0d_{5/2})^2$. 
Irrespective of such large differences in the configurations, the dineutron correlation is enhanced in the similar density region corresponding to $0.2<k<0.4~$fm$^{-1}$. 
This shows that the
dineutron phenomena may have a universal feature for two-neutron halo nuclei in that it appears strongly near the surface of the core.

\subsection{Theoretical aspects of dineutron in halo nuclei}\label{sec:ditheo}

\begin{figure}[htb]
 \begin{center}
   \includegraphics[width=7cm]{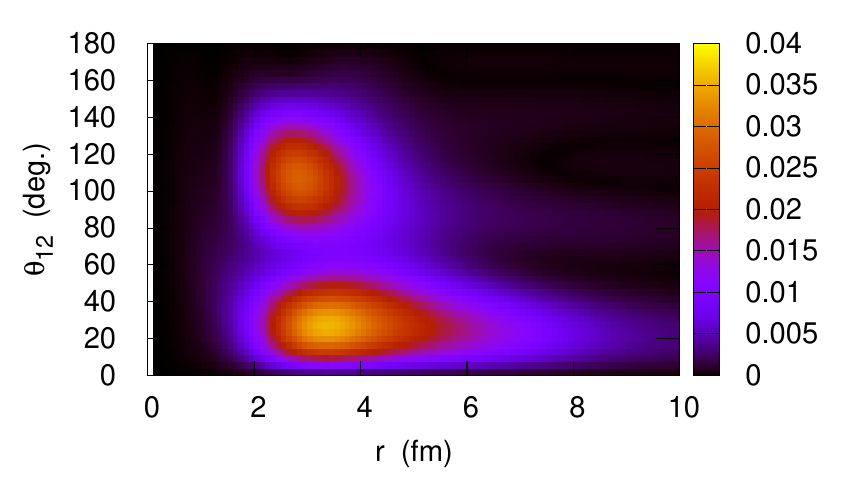}
   \includegraphics[width=7cm]{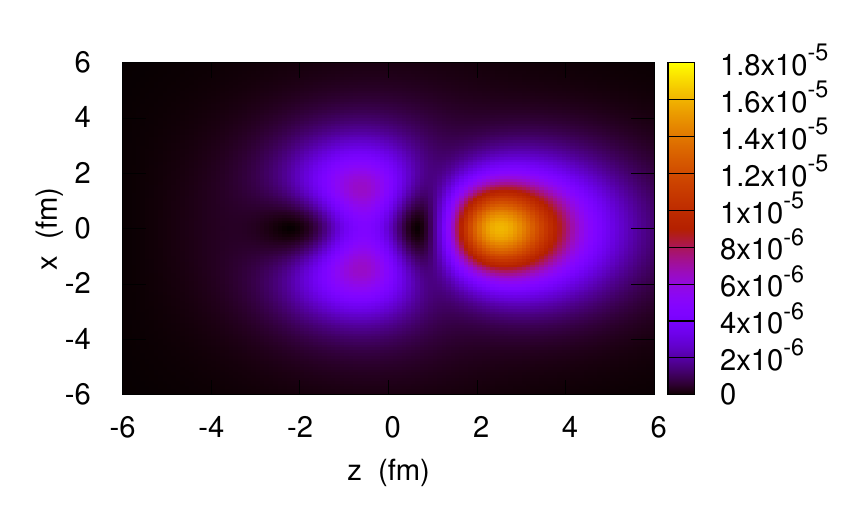}
   \caption{
   The two-particle density of $^{11}$Li obtained with the three-body model of Ref. \cite{HASA05}. 
   The upper panel plots the density as a function of $r_1=r_2=r$ and the angle $\theta_{12}$ between $\vec{r}_1$ and $\vec{r}_2$. 
   The weight factor of $8\pi^2r^4\sin\theta_{12}$ has been multiplied. The 
   lower panel shows the density of the first neutron when the second neutron is put on the $z$-axis at $z=3.4$ fm 
   }
   \label{fig:11Lidensity}
   \end{center}
\end{figure}

The dineutron correlation has been studied with three-body models \cite{BEES91,ZHUK93,OGAN99,HASA05}, 
the Hartree-Fock Bogoliubov theory \cite{MATS05,PILL07}, 
and the anti-symmetrized molecular dynamics (AMD) \cite{KOBA11}. 
As an example, Fig. \ref{fig:11Lidensity} shows the two-particle density for $^{11}$Li obtained with the 
three-body model of Ref. \cite{HASA05}. The upper panel of the figure shows the two-particle density $\rho(\vec{r}_1,\vec{r}_2)$ with $|\vec{r}_1|=|\vec{r}_2|\equiv r$, where $\vec{r}_1$ and $\vec{r}_2$ are the coordinates of the 
valence neutrons relative to the core nucleus, plotted as a function of $r$ and  
the angle $\theta_{12}$ between $\vec{r}_1$ and $\vec{r}_2$. The weight factor of $8\pi^2r^4\sin\theta_{12}$ has been multiplied to the density. One can clearly see that the peak around 
$\theta_{12}=30$ deg. (the `dineutron' component) 
is enhanced compared to the peak around $\theta_{12}=110$ deg. (the `cigar-like' component), see Fig. \ref{fig:correlation}.
On the other hand, the lower panel of Fig. \ref{fig:11Lidensity} shows the density distribution 
of the first neutron when the second neutron is placed on the $z$-axis at $z=3.4$ fm. One can see that the nearside 
configuration in which the first neutron is located close to the second neutron, $\vec{r}_1\sim\vec{r}_2$, 
is enhanced, while the farside component with $\vec{r}_1\sim-\vec{r}_2$ is largely suppressed. 
These are characteristic features of the dineutron correlation. 

\begin{figure}[tb]
 \begin{center}
   \includegraphics[width=10.cm]{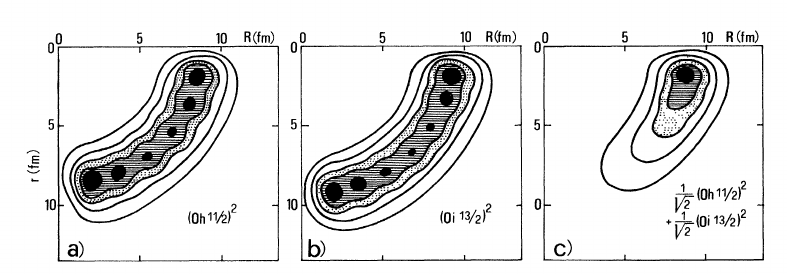}
   \caption{
   Two-particle density as a function of $\vec{R}=(\vec{r}_1+\vec{r}_2)/2$ and $\vec{r}=\vec{r}_1-\vec{r}_2$, 
   where $\vec{r}_1$ and $\vec{r}_2$ are the coordinates of two valence neutrons with respect to the core. 
   The right, the middle, and the left panels are obtained assuming $(0h_{11/2})^2$, 
   $(0i_{13/2})^2$, and $(0h_{11/2})^2+(0i_{13/2})^2$ configurations, respectively, with harmonic oscillator 
   wave functions. Taken from Ref. \cite{CATA84} 
   }
   \label{fig:catara}
   \end{center}
\end{figure}

In the three-body models, the wave function of the ground state with the spin and parity of $I^\pi=0^+$ is 
given by, 
\begin{equation}
\Psi(\vec{r}_1,\vec{r}_2)={\cal A}\sum_{n,n',j,l}C_{nn'jl}[\phi_{njl}(\vec{r}_1)\phi_{n'jl}(\vec{r}_2)]^{(00)},
\label{eq:wf2}
\end{equation}
with 
\begin{equation}
[\phi_{njl}(\vec{r}_1)\phi_{n'jl}(\vec{r}_2)]^{(00)}
=\sum_m\langle jmj-m|00\rangle\phi_{njlm}(\vec{r}_1)\phi_{n'jl-m}(\vec{r}_2),
\end{equation}
where ${\cal A}$ is the antisymmetrizer and
$\phi_{njlm}(\vec{r})$ is a single-particle wave function with the orbital angular momentum $l$, 
the total angular momentum $j$ and its $z$-component $m$, and the radial quantum number $n$. 
Here, the two single-particle wave functions are coupled to the 
total angular momentum $I$ and its $z$-component $I_z$ of $I=I_z=0$. $C_{nn'jl}$ is an expansion coefficient. 
An essential origin of the dineutron correlation is the admixture of different parity configurations, specifically of 
even-$l$ and odd-$l$ components, as demonstrated in Ref.~\cite{CATA84}. 
Figure \ref{fig:catara} shows the two-particle density as a function of the 
center of mass coordinate of the two neutrons, $\vec{R}=(\vec{r}_1+\vec{r}_2)/2$, and the 
relative coordinate, $\vec{r}=\vec{r}_1-\vec{r}_2$, 
obtained with harmonic oscillator wave functions for 
$\phi_{njlm}(\vec{r})$ \cite{CATA84}. The left and middle panels correspond to the pure $(0h_{11/2})^2$ 
and $(0i_{13/2})^2$ configurations, respectively. One can see that the density is significantly 
extended for both the $R$ and the $r$ coordinates. On the other hand, if these configurations are linearly superposed 
with an equal weight, the density distribution is well confined within a small value of $r$ (see 
the rightmost panel), 
%{\it that} 
which is nothing but the dineutron correlation. 

This feature can easily be understood if one symbolically writes Eq. (\ref{eq:wf2}) as 
\begin{equation}
\Psi(\vec{r}_1,\vec{r}_2)=\alpha \Psi_{\rm ee}(\vec{r}_1,\vec{r}_2)+\beta \Psi_{\rm oo}(\vec{r}_1,\vec{r}_2), 
\label{eq:3-body-oddeven}
\end{equation}
where $\Psi_{\rm ee}$ and $\Psi_{\rm oo}$ are the two-particle wave functions constructed with even-partiy and odd-parity states, respectively. 
The two-particle density $\rho(\vec{r}_1,\vec{r}_2)=|\Psi(\vec{r}_1,\vec{r}_2)|^2$ then reads 
\begin{eqnarray}
\rho(\vec{r}_1,\vec{r}_2)&=&
|\alpha|^2 |\Psi_{\rm ee}(\vec{r}_1,\vec{r}_2)|^2+|\beta|^2 |\Psi_{\rm oo}(\vec{r}_1,\vec{r}_2)|^2 \nonumber \\
&&+2Re[\alpha^*\beta\Psi^*_{\rm ee}(\vec{r}_1,\vec{r}_2)\Psi_{\rm oo}(\vec{r}_1,\vec{r}_2)], 
\end{eqnarray}
where $Re$ means the real part. When $\vec{r}_2$ changes the sign, that is, $\vec{r}_2\to-\vec{r}_2$, 
by definition $\Psi_{\rm oo}$ changes the sign while $\Psi_{\rm ee}$ remains the same, that is, 
$\Psi_{\rm ee}(\vec{r}_1,-\vec{r}_2)=\Psi_{\rm ee}(\vec{r}_1,\vec{r}_2)$ and 
$\Psi_{\rm oo}(\vec{r}_1,-\vec{r}_2)=-\Psi_{\rm ee}(\vec{r}_1,\vec{r}_2)$. By setting $\vec{r}_1=\vec{r}_2=\vec{r}$, one 
therefore finds,
\begin{equation}
\rho(\vec{r},\vec{r})=
|\alpha|^2 |\Psi_{\rm ee}(\vec{r},\vec{r})|^2+|\beta|^2 |\Psi_{\rm oo}(\vec{r},\vec{r})|^2 
+2Re[\alpha^*\beta\Psi^*_{\rm ee}(\vec{r},\vec{r})\Psi_{\rm oo}(\vec{r},\vec{r})], 
\label{eq:rho2-1}
\end{equation}
and
\begin{equation}
\rho(\vec{r},-\vec{r})=
|\alpha|^2 |\Psi_{\rm ee}(\vec{r},\vec{r})|^2+|\beta|^2 |\Psi_{\rm oo}(\vec{r},\vec{r})|^2 
-2Re[\alpha^*\beta\Psi^*_{\rm ee}(\vec{r},\vec{r})\Psi_{\rm oo}(\vec{r},\vec{r})], 
\label{eq:rho2-2}
\end{equation}
Notice that Eqs. (\ref{eq:rho2-1}) and (\ref{eq:rho2-2}) differ only by the interference term. 
For an attractive interaction, the coefficients $\alpha$ and $\beta$ are such that the interference 
term in Eq. (\ref{eq:rho2-1}) is positive, leading to an enhancement of the nearside density, $\rho(\vec{r},\vec{r})$.
At the same time, the interference term suppresses the farside density, $\rho(\vec{r},-\vec{r})$. 
This is exactly what is seen in the dineutron correlation (see the lower panel of Fig. \ref{fig:11Lidensity}). 

\begin{figure}[tb]
 \begin{center}
   \includegraphics[width=4cm]{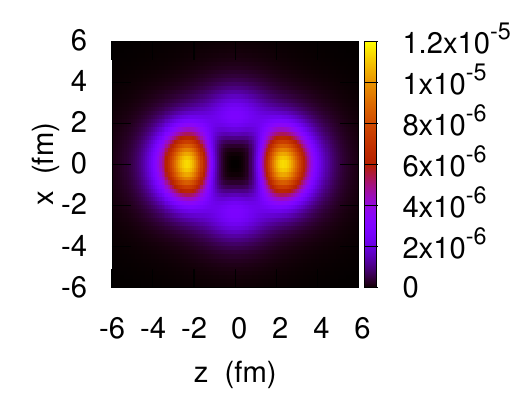}
    \includegraphics[width=4cm]{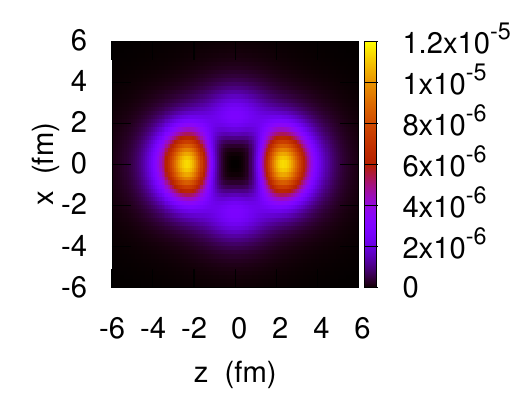}
     \includegraphics[width=4cm]{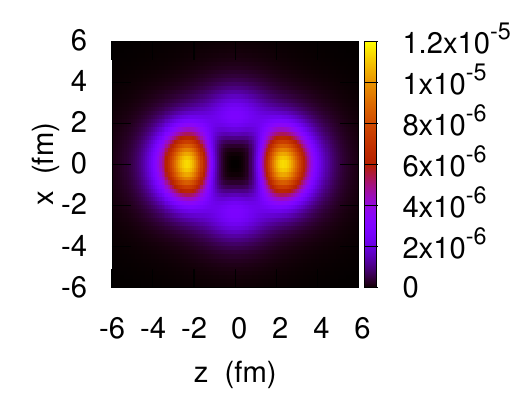}
   \caption{
   The density distribution of the first valence neutron in $^{18}$O when the second valence 
   neutron is located at 
   $z=1$ fm (the left panel), 2 fm (the middle panel), and 3 fm (the right panel). 
   These are obtained by assuming the pure (1$d_{5/2})^2$ configuration
}
   \label{fig:18o-nocorr}
   \end{center}
\end{figure}

\begin{figure}[tb]
 \begin{center}
   \includegraphics[width=4cm]{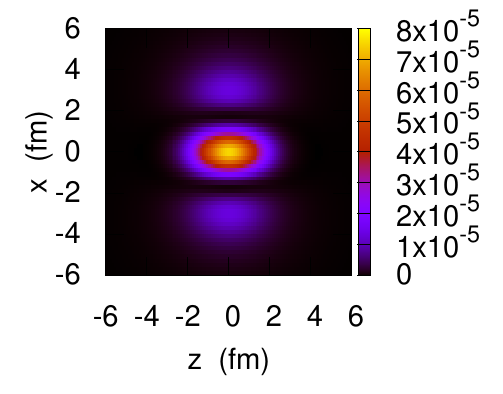}
    \includegraphics[width=4cm]{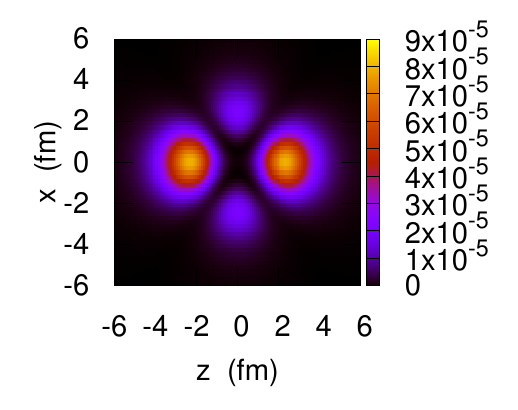}
     \includegraphics[width=4cm]{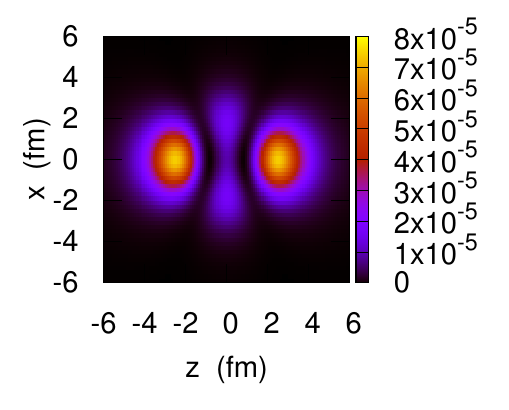}
   \caption{
   Same as Fig. \ref{fig:18o-nocorr}, but with the pairing correlation within the bound single-particle 
   states, 1$d_{5/2}$, 2$s_{1/2}$, and 1$d_{3/2}$ 
   }
   \label{fig:18o-nodineutron}
   \end{center}
\end{figure}

\begin{figure}[tb]
 \begin{center}
   \includegraphics[width=4cm]{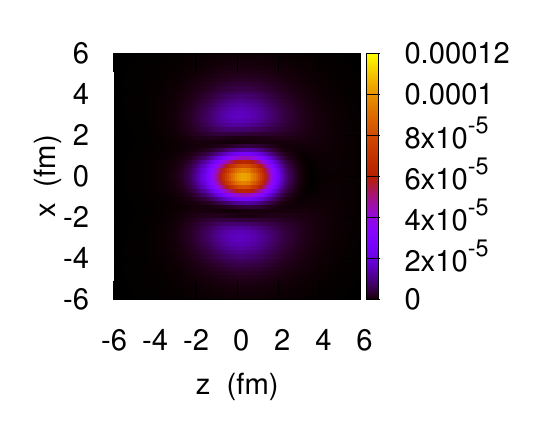}
    \includegraphics[width=4cm]{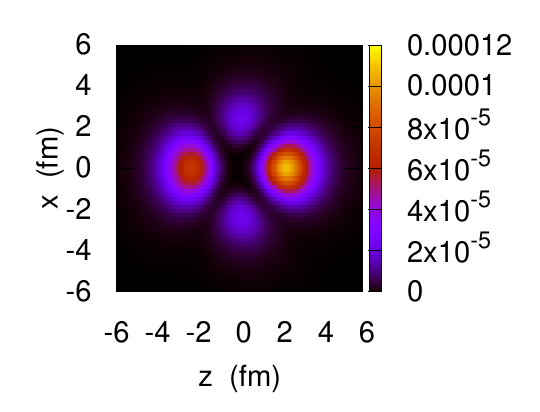}
     \includegraphics[width=4cm]{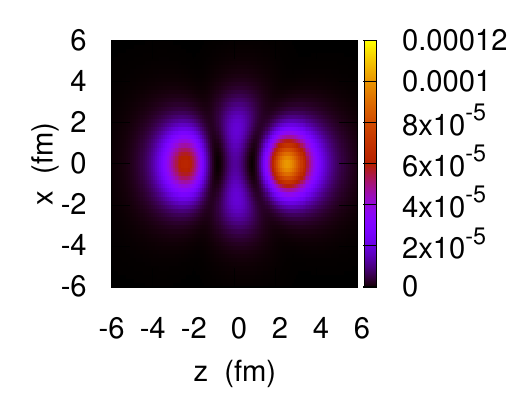}
   \caption{
   Same as Fig. \ref{fig:18o-nodineutron}, but with the pairing correlation within the bound single-particle 
   states, 1$d_{5/2}$, 2$s_{1/2}$, and 1$d_{3/2}$, as well as continuum single-particle states up to $l$=7 
   }
   \label{fig:18o-dineutron}
   \end{center}
\end{figure}

Let us demonstrate the role of mixing of different parity configurations in a different way. 
For this purpose, let us consider the ground state of $^{18}$O with the three-body model 
assuming the $^{16}$O+$n$+$n$ structure. The two-body subsystem, $^{16}$O+$n$, has three bound states, 
1$d_{5/2}$, 2$s_{1/2}$, and 1$d_{3/2}$. When there is no interaction between the two valence neutrons, those 
neutrons occupy solely the 1$d_{5/2}$ state. Fig. \ref{fig:18o-nocorr} shows the density distribution of the 
first valence neutron when the second neutron is put at $z_2$ on the $z$-axis. The left, the middle, and the right 
panels show the density for $z_2$=1, 2, and 3 fm, respectively. Since this case corresponds to the independent particle 
approximation, the first valence neutron does not care where the second valence neutron is, and thus the 
three density distributions are identical to each other. 
Fig. \ref{fig:18o-nodineutron} shows similar density distributions, but in the presence of the pairing correlation 
between the valence neutrons. Here, the pairing active space is restricted to the bound single-particle states, 
1$d_{5/2}$, 2$s_{1/2}$, and 1$d_{3/2}$. In this case, the two neutrons are correlated, and the density distribution 
is altered depending on the location of the second valence neutron. However, the single-particle states in the 
pairing active space are all even-parity states, and the peak on the nearside is identical to the peak on the farside. 
That is, in this case, there is only the first term in Eqs. (\ref{eq:rho2-1}) and (\ref{eq:rho2-2}), and 
there is no dineutron correlation there, even though the pairing correlation still exists. 
When the active space of pairing is extended to continuum states, the density distributions turn to those shown in 
Fig. \ref{fig:18o-dineutron}. The pairing active space includes both even-parity and odd-parity states, and 
one can now clearly see that the nearside component is significantly enhanced compared to the farside component, 
manifesting the dineutron correlation. 

As demonstrated in the example of $^{18}$O, as well as of $^{120}$Sn discussed in Ref. \cite{PILL07}, 
the dineutron correlation is not necessarily associated with weakly bound nuclei. However, one can 
expect that the dineutron correlation is enhanced in weakly bound nuclei mainly for the following two 
reasons. Firstly, weakly bound nuclei exhibit an extended density distribution. As the 
pairing gap in infinite nuclear matter has a peak at a density 
lower than the normal density \cite{MATS06}, the pairing correlation is effectively enhanced in the low-density region in weakly bound nuclei. 
Secondly, weakly bound nuclei have a low threshold to continuum spectrum, 
and thus the mixing of opposite parity states is much easier than in stable nuclei.

\begin{figure}[tb]
 \begin{center}
   \includegraphics[width=7.cm]{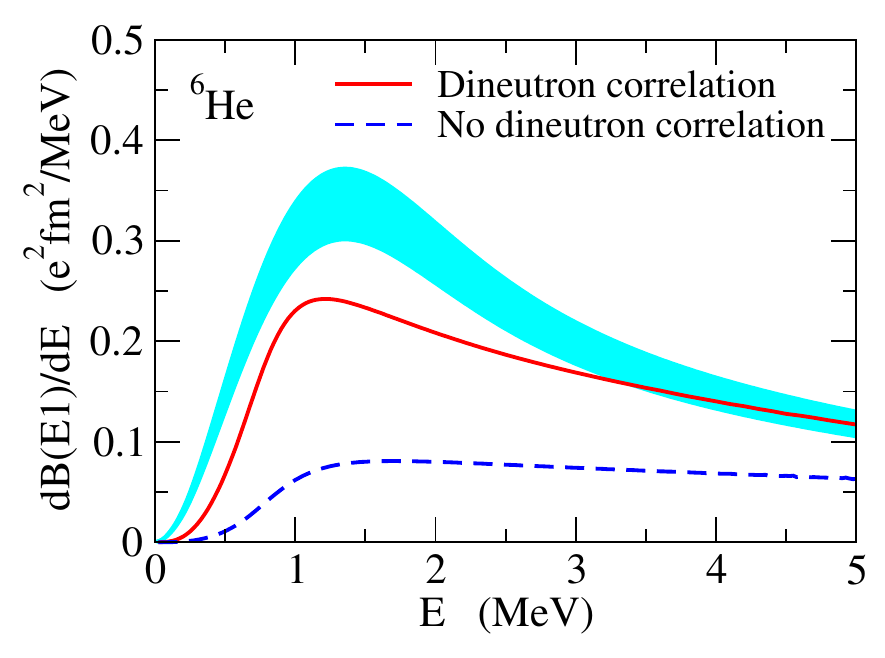}
    \includegraphics[width=5.cm]{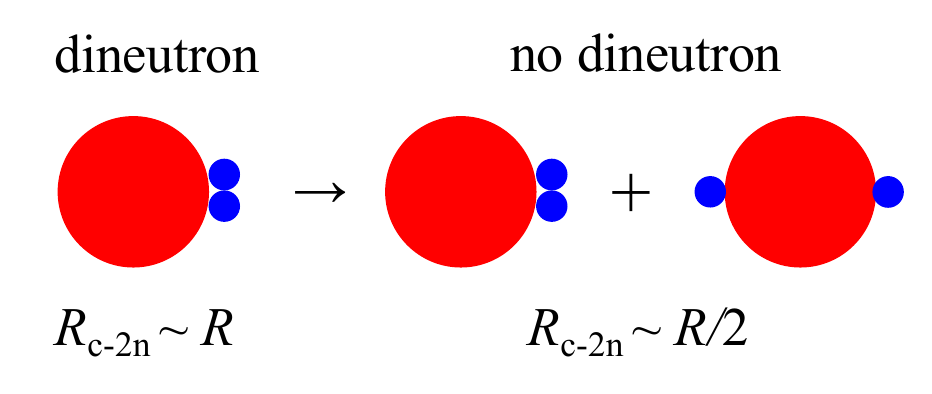}
   \caption{
   The $B(E1)$ distribution of $^6$He computed with the three-body model \cite{HASA09}. 
   The solid line shows the full response, while the dashed line is obtained by excluding even partial waves from the 
   ground state wave function so that the dineutron correlation disappears. 
   The shaded area denotes the experimental data \cite{SUN21}, which is also shown in Fig.~\ref{fig:beone6he}.
   The figure also provides a schematic 
   illustration of how the dineutron correlation alters the mean distance between the core nucleus and the 
   center of mass of the dineutron. Here, $R$ denotes the radius of the core nucleus 
      }
   \label{fig:6he-be1-3body}
   \end{center}
\end{figure}

The dineutron correlation also affects the electromagnetic transition of the halo nuclei. 
To demonstrate this, Fig. \ref{fig:6he-be1-3body} shows the $B(E1)$ strength of $^6$He obtained with 
the three-body model of Ref. \cite{HASA09}. The solid line shows the full strength, which takes into 
account the dineutron correlation in the ground state.
 The dashed line is obtained by 
including only odd-angular momentum states in the ground state. As has been shown, the dineutron correlation 
vanishes in this case. One can clearly see that the $B(E1)$ strength distribution is considerably reduced in 
the absence of the dineutron correlation. 
The reduction of the $B(E1)$ distribution can be easily understood using the cluster sum rule, 
Eq. (\ref{eq:eonesum2}). 
In the pure dineutron configuration, the two valence neutrons are located close to each other around the surface of the core nucleus, and the mean distance between the core nucleus and the center of mass of the dineutron, $r_{c-nn}$, 
is similar to the radius of the core nucleus, $R$ \footnote{In reality, 
$r_{c-nn}$ should be somewhat larger than $R$ since the halo wave function is extended from the core nucleus.}. 
On the other hand, in the absence of the dineutron correlation, the dineutron and the cigar-like configurations 
equally contribute to the ground state wave function. Since $r_{c-nn}$ is zero for the cigar-like configuration, the mean distance becomes about half of $R$ in the absence of dineutron correlation, resulting in a reduction 
in the distribution of $B(E1)$. 
In the actual calculation of $^6$He, the value of $\sqrt{\langle r^2_{c-nn}\rangle}$  is reduced 
from 3.63 fm \footnote{This value can be compared to 
the empirical values, $\sqrt{\langle r^2_{c-nn}\rangle}$ =3.9(2) fm from the 
Coulomb excitations \cite{SUN21} and 3.71(07) fm from the charge radius measurement \cite{BERT07}.}
to 2.61 fm when the dineutron correlation is switched off. Notice that the actual ground state wave function contains a small fraction 
of the cigar-like configuration, and thus the scaling deviates from that for the pure dineutron configuration.

\section{Dineutron in unbound nuclei}\label{sec:unbound}

%In not only the bound nuclei, such as two-neutron halo nuclei, but also the unbound nuclei, we may find a resonance with a dineutron correlation, which is discussed here. 
We expect that $nn$ correlations become stronger when the absolute value of the neutron separation
energy is well below the pairing gap, $|S_{n}|\ll \Delta$, as pointed out by Dobaczewski et al.~\cite{DOBA07}.
Such consideration suggests that not only very weakly bound nuclei, such as two-neutron halo nuclei discussed above, but also a barely unbound two-neutron emitter should have a strong $nn$ correlation.
We are in particular interested in the dineutron correlation: the spatially compact $nn$, as in the two-neutron halo nuclei.

\begin{figure}[h]
 \begin{center}
   \includegraphics[width=\linewidth]{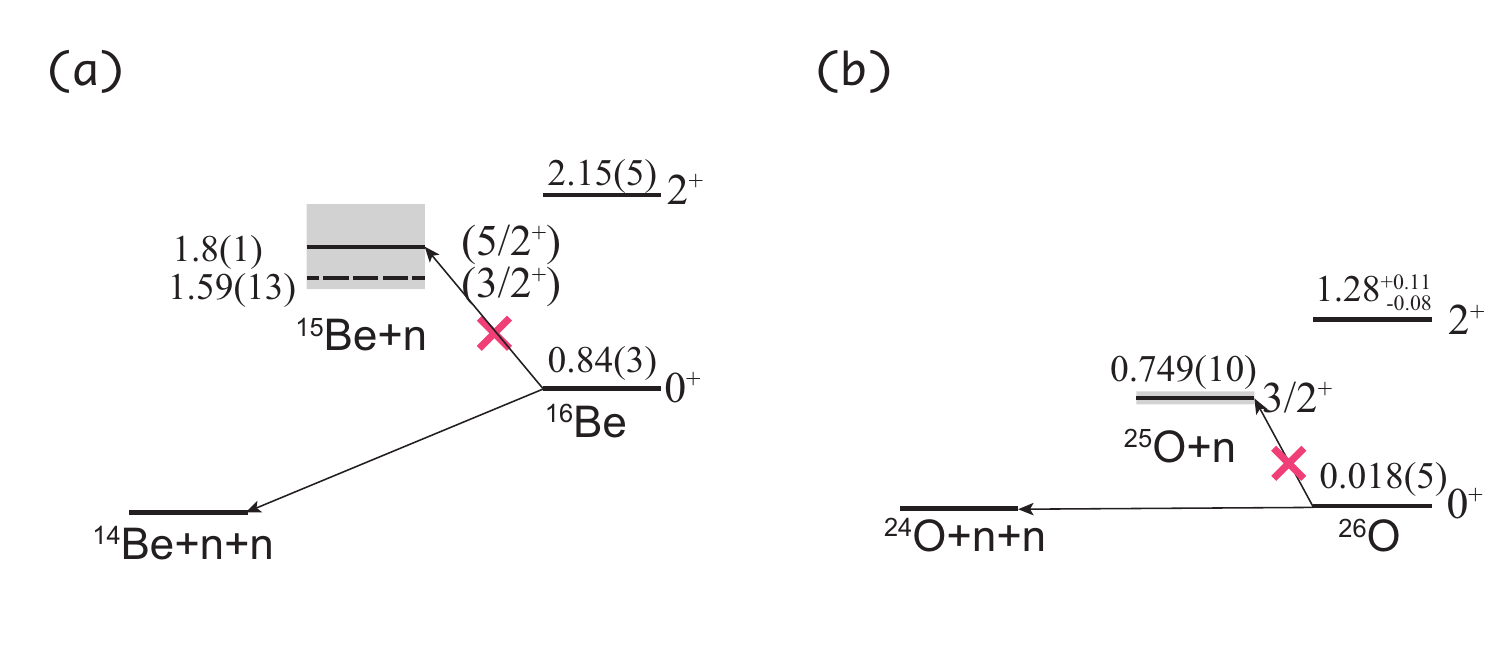}
   \caption{Candidate states with dineutron correlations in low-energy two-neutron emitters: (a) $^{16}$Be~\cite{MONT24}, and (b) $^{26}$O~\cite{KOND16}, 
   where energies are shown in MeV relative to the energy level of $^{14}$Be+$n$+$n$ and $^{24}$O+$n$+$n$, respectively.
   The known energy levels of $^{15}$Be and $^{25}$O are also shown.
   The 3/2$^+$ state in $^{15}$B can be the ground state, which was observed only in the sequential decay via the $^{14}$Be($2_1^+$) state, followed by the decay into $^{12}$Be+$n$+$n$ ($3n$ emission channel)\cite{KUCH24}, which was not discussed in Ref.~\cite{MONT24}. The experimental widths for the 5/2$^+$ state in $^{15}$B~\cite{SNYD13} and the ground state of $^{25}$O~\cite{KOND16} are shown by the shaded region}
   \label{fig:threebodydecay}
   \end{center}
\end{figure}

Promising dineutron candidates in unbound systems are two-neutron emitters, such as $^{16}$Be and $^{26}$O. 
%The unbound excited states in the bound nuclei may also show the dineutron correlation, as in the case of the first excited state of $^{6}$He: $^6$He($2_1^+$). 
The energy levels of these cases are shown in Fig.~\ref{fig:threebodydecay}.
In the following, we first discuss recent results and near-future perspectives for the three-body decay experiments on the two neutron emitters: $^{16}$Be and $^{26}$O.
Theoretical aspects of the dineutron correlation in unbound nuclei are also discussed.  

\newcommand{\efnn}{E_{\rm rel}}

\subsection{Three-body decay experiments for unbound nuclei}\label{sec:unboundexp}

Dineutron-correlation experiments for unbound nuclei are still scarce.
Here, we focus on recent experiments for $^{16}$Be~\cite{MONT24} and $^{26}$O~\cite{KOHL15}.
As shown, while we have a hint of dineutron in the ground state of $^{16}$Be~\cite{MONT24}, the experiment on $^{26}$O was inconclusive~\cite{KOHL15}.

\subsubsection{Three-body decay of $^{16}$Be}
\label{sec:16Be}

$^{16}$Be is an unbound nucleus, two neutrons beyond the last bound berylium isotope, $^{14}$Be. The pioneering experiment in $^{16}$Be was performed at MSU
using a one-proton removal reaction of $^{17}$B 
at 53 MeV/u~\cite{SPYR12}. This experiment 
observed the candidate ground state lying at 1.35(10) MeV above the
$^{14}$Be+$n$+$n$ threshold, 
approximately 500~keV higher than the most recent observation ($\erel=$0.84(3) MeV) shown in 
Fig.~\ref{fig:threebodydecay}(a)~\cite{MONT24}.
The $^{16}$Be ground state was found to be a two-neutron emitter as shown below.
%, since no decay branch was observed through $^{15}$Be. 

The observation of $^{15}$Be is crucial for determining whether $^{16}$Be decays via direct two-neutron emission or sequentially through an intermediate $^{15}$Be state. 
The current understanding of $^{15}$Be is shown in Fig~\ref{fig:threebodydecay}(a). A resonance with $E_r=$1.8(1) MeV and $\Gamma=0.58(20)$~MeV was observed in the $^{14}$Be($d,p$) reaction at 59 MeV/u at MSU~\cite{SNYD13}, in the $^{14}$Be+$n$ decay channel. 
More recently, an experiment using the multi-nucleon removal reaction from $^{18}$C at SAMURAI at RIBF observed a $^{15}$Be state at $\erel=1.70(13)$~MeV~\cite{CORS21}, consistent with 
Ref.~\cite{SNYD13}.
This $(d,p)$ experiment tentatively assigned a spin-parity of 
5/2$^+$ to the 1.8 MeV state, as the shell-model calculation combined with reaction-model calculation indicates that 
the 5/2$^+$ state predominantly decays into $^{14}$Be$_{gs}$+$n$, whereas the nearby 3/2$^+$ state is expected to decay mainly via the first excited state of $^{14}$Be, $^{14}$Be(2$_1^+$). 
The $^{14}$Be(2$_1^+$) state lies at $E_x=1.54(13)$~MeV~\cite{SUGI07}, and is unbound 
with respect to two-neutron emission. This implies that the 3/2$^+$ state of $^{15}$Be leads to the $^{12}$Be+$n$+$n$+$n$ channel.
Indeed, a $3n$-coincidence analysis of the same data set revealed a state at $E_r=1.59(13)$~MeV 
(defined relative to the $^{14}$Be+$n$ threshold), as the candidate for the $3/2^+$ state~\cite{KUCH24}. 
Since the $^{16}$Be experiment at RIBF discussed here measures the $^{14}$Be+$n$+$n$ channel~\cite{MONT24}, we restrict our discussions to the effect of the 1.8 MeV state in $^{15}$Be.

The sequential decay of $^{16}$Be via the 1.8 MeV state in $^{15}$Be should be negligible, as this process is largely energetically forbidden. Nevertheless, we assess the possible sequential decay width arising from the finite width ($\Gamma=0.58(20)$ MeV) of the intermediate $^{15}$Be($5/2^+$) state.
Using the formalism presented in~Refs.~\cite{VOLY06,VOLY12}, we evaluate the sequential decay width to be about 0.5 keV, which is more than two orders of magnitude smaller than the observed width of the $^{16}$Be ground state
($\Gamma=0.32$~MeV). This confirms that 
the sequential decay contribution is indeed negligible.

%\red{We briefly review the observation of $^{15}$Be before going into $^{16}$Be.}
%The first attempt to produce $^{15}$Be was made
%at MSU using the two-proton knockout reaction of $^{17}$C 
%at 55 MeV/u~\cite{SPYR11}, which did not show any peaks for $^{15}$Be in the invariant spectrum $^{14}$Be +$n$, indicating that the possible $^{15}$Be state 
%decays through the state of $^{14}$Be(2$_1^+$) at $E_x=1.54(13)$~MeV~\cite{SUGI07}, which is unbound with respect to a two-neutron emission. The state of $^{15}$Be was not observed because it then emits three neutrons, which were not detectable at that time. 
%The following experiment at MSU used the $^{14}$Be($d,p$)$^{15}$Be reaction, which showed a single peak at $\erel=1.8(1)$~MeV in the $^{14}$Be+$n$ channel and was tentatively assigned as 5/2$^+$~\cite{SNYD13}. 

%The three-neutron decay channel for the same $(d,p)$ data~\cite{SNYD13} was
%recently analyzed, and the four-body decay energy $E_{fnnn}$($^{12}$Be-$n$-$n$-$n$) showed a peak at  $E_{fnnn}=0.330(20)$~MeV, which corresponds to the energy relative to $^{14}$Be+$n$ being 1.59(13)~MeV~\cite{KUCH24} (See Fig.\ref{fig:threebodydecay}(a)). This state is tentatively assigned as 3/2$^+$. We note that the shell model predicts the nearby states 
%5/2$^+$ and 3/2$^+$ for the ground and the first excited state of $^{15}$Be.
%It is now clear that the ground state of $^{15}$Be is located at rather high energies relative to $^{14}$Be+$n$,
%which prohibits $^{16}$Be from decaying through $^{15}$Be. 

\begin{figure}[h]
 \begin{center}
   \includegraphics[width=8.cm]{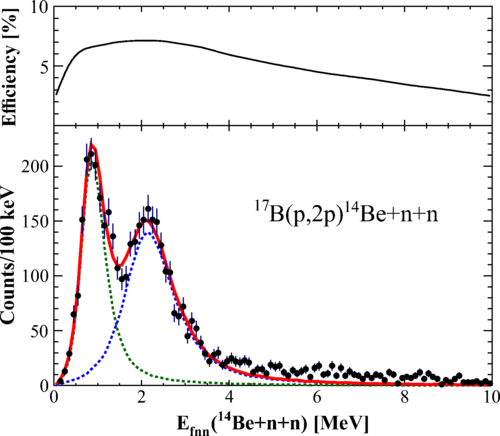}
   \caption{The three-body decay energy spectrum of $^{16}$Be produced by the $^{17}$B($p,2p$) reaction at 277 MeV/u at RIBF, RIKEN. The upper panel shows the efficiency curve. The figure is adopted from Ref.~\cite{MONT24}}
   \label{fig:be16}
   \end{center}
\end{figure}

\begin{figure}[h]
 \begin{center}
   \includegraphics[width=9.cm]{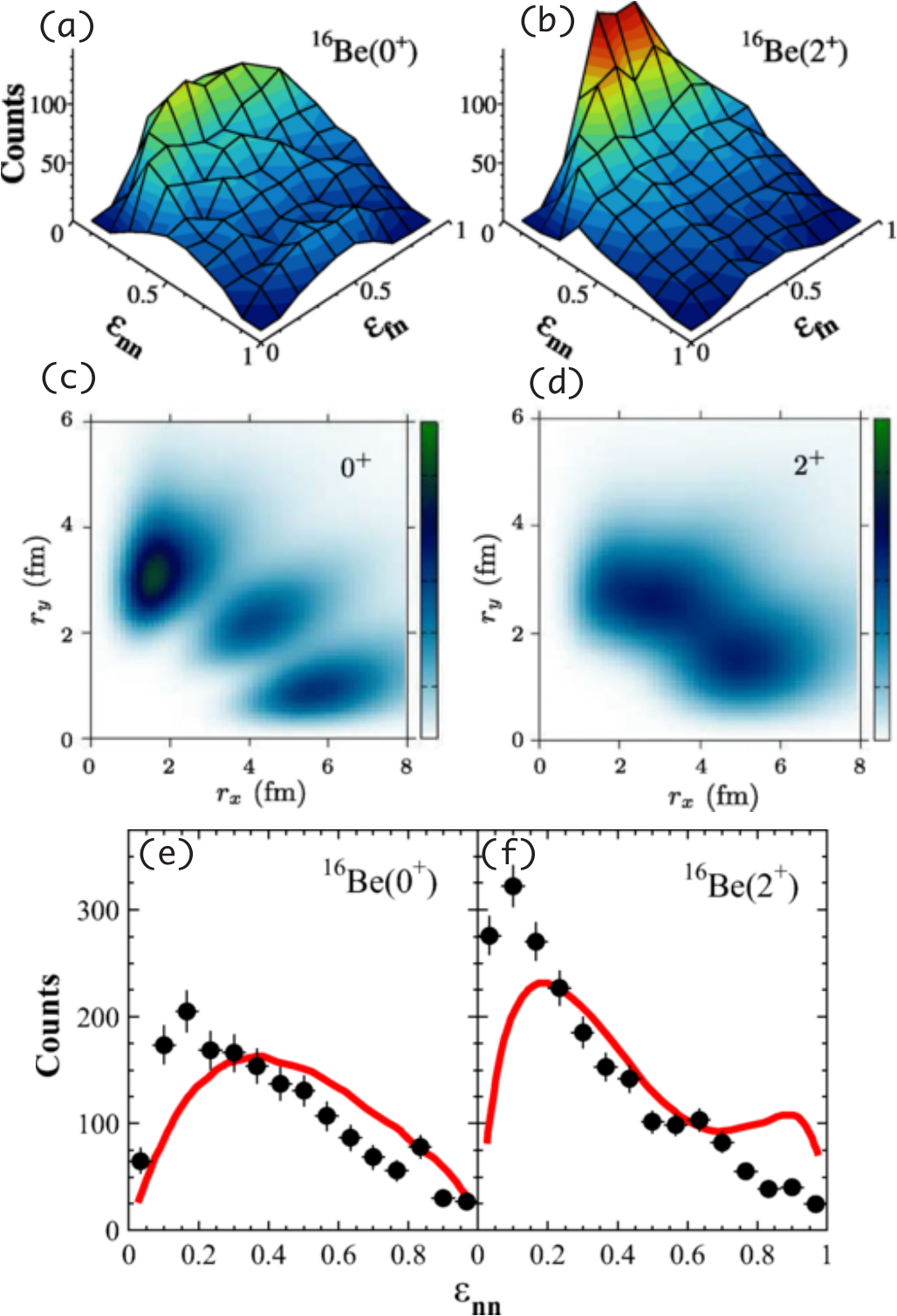}
   \caption{(a)(b): Dalitz plots of the normalized two-body energies, $\varepsilon_{fn}$($^{14}$Be-$n$) vs. $\varepsilon_{nn}$ ($n$-$n$) for the three-body decay of the $^{16}$Be ground state (a) and the first excited state (b). (c)(d): Spatial probability distributions,$P(r_x,r_y)$, for these two states calculated within the three-body theory. The probability is shown as a function of the distance $r_x$($n$-$n$) and $r_y$($^{14}$Be-$nn$).
   (e)(f): Normalized $nn$ relative energy spectra for
   the ground (e) and first excited states (f),
   compared with theoretical results corresponding to the two-neutron density distributions shown in (c) and (d), respectively.
   Overall agreement between the data and the theory is obtained for both states.
   The figure is adapted from Ref.\cite{MONT24}}
   \label{fig:be16cor}
   \end{center}
\end{figure}

The earlier experiment on $^{16}$Be~\cite{SPYR12} observed the narrow 
opening angle of the outgoing two neutrons in the forward direction, for which they claimed discovery of the dineutron correlation. However, we should note that this correlation is inherent in final-state interactions (FSI) of the two neutrons, not in the spatially-compact dineutron lying in the ground state of $^{16}$Be, as pointed out by Ref.~\cite{MARQ12}. We should distinguish between the two concepts: dineutron in the initial state and that in the final breakup states. For the final state interactions of the two neutrons, due to the large $nn$ scattering length in magnitude for low relative energies, it naturally favors the narrow angles.
 On the other hand, if the spatially compact dineutron correlation occurs in the initial state, the relative momentum between the two neutrons tends to be larger due to the uncertainty principle of quantum mechanics. We thus consider that a large opening angle in the momentum space, which corresponds to back-to-back emission of two neutrons in the c.m. of the three-body resonance, can be a signature of the dineutron. Such a back-to-back emission is theoretically predicted for the decay of the $^6$He(2$_1^+$) state into $^4$He+n+n~\cite{KIKU13}.
 We also show later in Fig.~\ref{fig:26o-density-rp} that the opening angle in momentum space is about 160 degrees in the decay
 of $^{26}$O. Note that the localized large relative momentum distribution may be relevant to the three-body system. For a simple two-body fermionic systems with compact size (like only $nn$ in free space), the momentum distribution is wider but its expectation value can be small.
  This is also similar to the case of the two-neutron halo nuclei: as shown above for the quasi-free scattering experiment of $^{11}$Li~\cite{KUBO20}, the larger opening angle in the momentum space was observed as evidence of the dineutron
 (See Sec.~\ref{sec:quasi}).

 At RIBF at RIKEN, $^{16}$Be states were populated via quasi-free proton scattering, $^{17}$B($p,2p$), at 277 MeV/nucleon with high statistics achieved using a very thick (15~cm) liquid hydrogen target system, MINOS~\cite{OBER11,SANT18}. The resulting spectrum is shown in Fig.~\ref{fig:be16}~\cite{MONT24}.
 This result is obtained from the same experiment as
 the quasi-free proton scattering of $^{11}$Li described in Sec.~\ref{sec:quasi}, where a cocktail beam of $^{11}$Li, 
 $^{14}$Be, and $^{17}$B was delivered to the same SAMURAI setup.
 The spectrum clearly exhibits two distinctive peaks in the three-body decay energy at 0.84(3)~MeV and 2.15(5)~MeV.
 The former is assigned to the ground state, and the latter to the first $2^+$ state. We infer that the earlier MSU result at $\efnn=1.35$~MeV may be a mixture of these two states due to a poorer energy resolution and much lower statistics.

 Figures \ref{fig:be16cor}(a) and (b) show the Dalitz plots for the three-body decay of the 
 $^{16}$Be ground state (a) and the first excited state (b) in terms of the normalized two-body energies, 
 $\varepsilon_{fn}=E_{fn}/E_{fnn}$ and $\varepsilon_{nn}=E_{nn}/E_{fnn}$. 
 $E_{fn}$ and $E_{nn}$ denote the two-body decay energies between $^{14}$Be and the neutron, and between the two neutrons, respectively. When there are no interactions among two-body systems, three-body decay proceeds via phase-space decay, resulting in a uniform distribution in the Dalitz plot. However, both figures show an enhancement at $\varepsilon_{nn}\sim 0.1$, 
 as can be seen more clearly in their projections
shown in Fig.\ref{fig:be16cor}(e)(f). These enhanced structures may arise from final-state interactions that favor low $E_{nn}$.

 Theoretical analysis using the three-body model, which also treats a realistic description of the 
 decay of $^{16}$Be~\cite{CASA18, CASA19}, elucidated the dineutron correlation, as shown in Fig.\ref{fig:be16cor}(c)(d). These figures show two-neutron probability densities,
 denoted by $P(r_x,r_y)$, for the ground state and the first 2$^+$ state, where $r_x$ and $r_y$ represent
 the distance between the two neutrons ($n$-$n$) and 
 between the $^{14}$Be core and the c.m. of the two neutrons ($^{14}$Be-$nn$), in the Jacobi $\bold{T}$ coordinate system, respectively.
 The calculation exhibits dineutron correlation in the ground state, while the compact dineutron component is small in the first excited state.

 The comparison of $\varepsilon_{nn}$ between the experimental data and the calculations of the three-body model is shown in Fig.\ref{fig:be16cor}(e)(f). 
 An overall agreement indicates that the dineutron correlation exists in the ground state of $^{16}$Be. 
 The peak at low $\varepsilon$ due to the FSI is more pronounced for
 the $2^+$ state, which may reflect the minor contribution of the compact dineutron in this state.
 
 There is some contribution near $\varepsilon_{nn}\sim 1$, which may be a reflection of the dineutron in the initial state. However, no clear enhancement is observed.
 The three-body decay dynamics may be more complex than the simple view that the compact dineutron favors back-to-back decay. More experimental and theoretical investigations are needed to further clarify dineutron correlations and three-body decay dynamics of the two-neutron emitters. For instance, in the three-body model calculation shown above, the $^{15}$Be(3/2$^+$) state is not properly included as the core excited state is not treated in the theory, which should be investigated.
 
 \subsubsection{Three-body decay of $^{26}$O}\label{sec:o26}

We expect a stronger $nn$ correlation for a nucleus with $|S_{2n}|\sim 0$ as mentioned in Ref.~\cite{DOBA07}. 
The ground-state of $^{26}$O lies 
only 18$\pm$3(stat.)$\pm$4(syst.)~keV above the decay threshold of $^{24}$O + $n$ + $n$, having 
the smallest two-neutron energy among known two-neutron emitters~\cite{KOND16}. This was observed by the one-proton removal reaction of $^{27}$F at 201~MeV/nucleon at SAMURAI at RIBF.
The $^{26}$O ground state is a pure two-neutron emitter, since it 
cannot decay into $^{25}$O$+n$ (even if the width of the $^{25}$O state, 0.088(6) MeV \cite{KOND16} is taken into account) but only into $^{24}$O$+n+n$ as shown in Fig.~\ref{fig:threebodydecay}(b).

\begin{figure}[ht]
 \begin{center}
   \includegraphics[width=\linewidth]{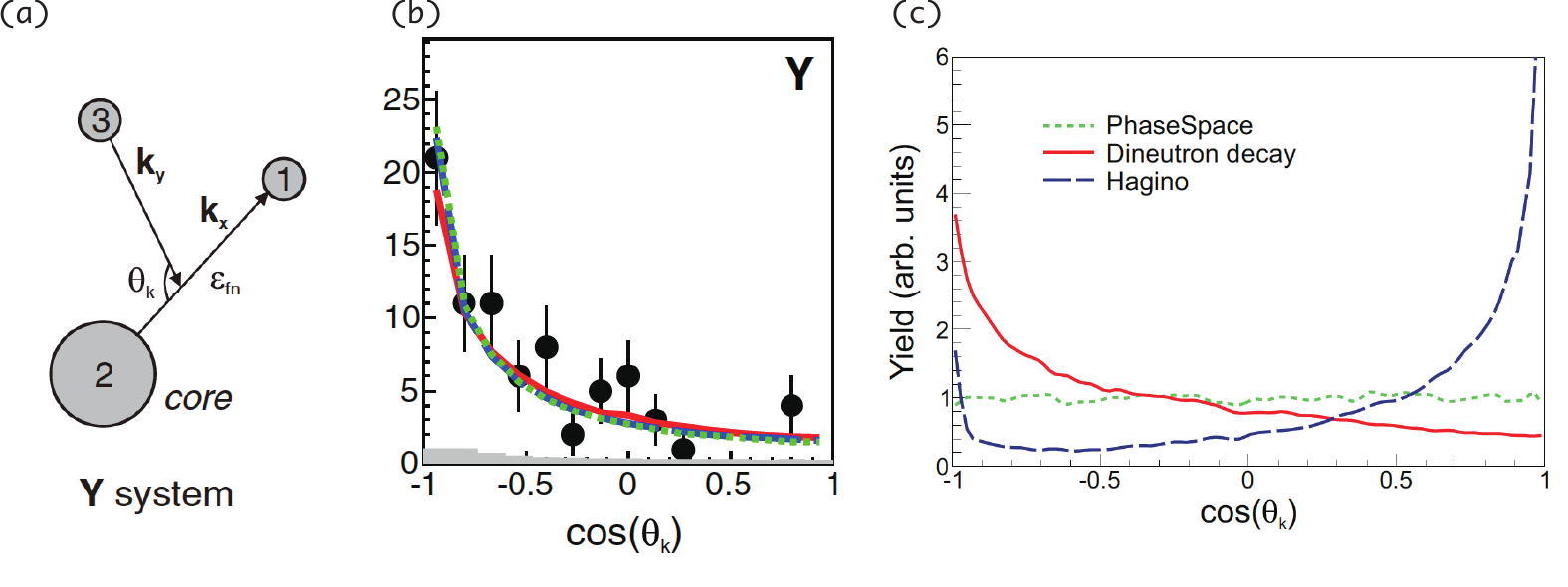}
   \caption{(a) Definition of the coordinate system (Jacobi $\bold{Y}$) and the correlation angle $\theta_k$; (b) experimental $\cos \theta_k$ distribution, where the red, blue, and green curves correspond to $nn$ FSI, 
   the dineutron (three-body model), and phase-space decay, respectively; (c)  theoretical predictions for $\cos \theta_k$, where the same color coding as in (b).
   The figure is adapted from Ref.~\cite{KOHL15}}
   \label{fig:kohley}
   \end{center}
\end{figure}

The existence of 
a spatially compact dineutron correlation in $^{26}$O
was predicted by Hagino and Sagawa and is discussed in more detail
in Section~\ref{sec:unboundtheo}.
They predict a characteristic angular correlation
between the two neutrons emitted from the ground state
of $^{26}$O, as shown in Fig.~\ref{fig:26o-angular} and Fig.~\ref{fig:kohley}(c).

%A pioneering experiment was performed at MSU by Z.Kohley
%using the MoNA neutron detector array\cite{KOHL15}. 

Kohley et al. performed a pioneering measurement 
of the angular correlation of the valence two neutrons in $^{26}$O
 at MSU, whose result is shown in Fig.~\ref{fig:kohley}~(b)~\cite{KOHL15}.
The state of $^{26}$O was populated by the one-proton removal of $^{27}$F at 82 MeV/nucleon.
The angular distribution characteristic of the 
dineutron correlation 
is predicted by the three-body model by Hagino and Sagawa~\cite{HAGI14}, where
two neutrons are emitted back-to-back due to the quantum mechanical uncertainty principle. 
The details of this mechanism are discussed below in Sec.~\ref{sec:unboundtheo}
(see also Figs. \ref{fig:26o-angular} and \ref{fig:26o-density-rp}).
That is, the relative momentum 
between the two neutrons tends to be larger, reflecting the localization of the two neutrons, which is
shown in the blue dashed curve (Hagino's dineutron) in Fig.\ref{fig:kohley}~(c).
Kohley has another model, which they call {\it dineutron decay}. However, this is {\it not} a dineutron in the current context of the compact, spatially correlated $nn$ system. In fact, this model reflects the final-state interaction (FSI) between $nn$ with a long scattering length, so we would rather call this $nn$ with FSI (``$nn$ FSI" for brevity). 
Two neutrons with $nn$ FSI tend to be emitted in the same direction (with low relative momentum).

The experiment in Ref.~\cite{KOHL15} failed to
distinguish these two possibilities shown by the blue (Hagino's dineutron) and red ($nn$ FSI) curves.
These curves match exactly the predicted phase-space decay 
due to the lack of angular resolutions~\cite{KOHL15}, as shown in Fig.\ref{fig:kohley} (b). For neutrons of about 80~MeV emitted from the resonance energy of only 18~keV, the maximum opening angle is only about 20~mrad. Since the neutron detector array MoNA, whose one module has dimensions of 10~$\times$~10~$\times$~200~cm$^3$~\cite{BAUM05}, 
was located at 6.05~m~\cite{LUND12}, the distance between the two neutrons in the perpendicular direction at this detector location should be about 12~cm at maximum, 
which is compatible with the detector width of 10~cm. 
Hence, the angular resolution was insufficient.
A position resolution of about 2~cm or less is necessary to extract the angular correlation. 

\subsubsection{High granularity neutron detector array: HIME}

The angular correlation measurement discussed in the previous Section requires a significantly higher position resolution for neutron detections.
At SAMURAI at RIKEN, we have developed a high-granularity neutron detector array, called HIME ({\bf HI}gh resolution detector array for {\bf M}ulti 
neutron {\bf E}vents). 
One of the main goals of building HIME 
is to experimentally extract the angular correlation of the ground state of $^{26}$O.

\begin{figure}[h]
 \begin{center}
   \includegraphics[width=\linewidth]{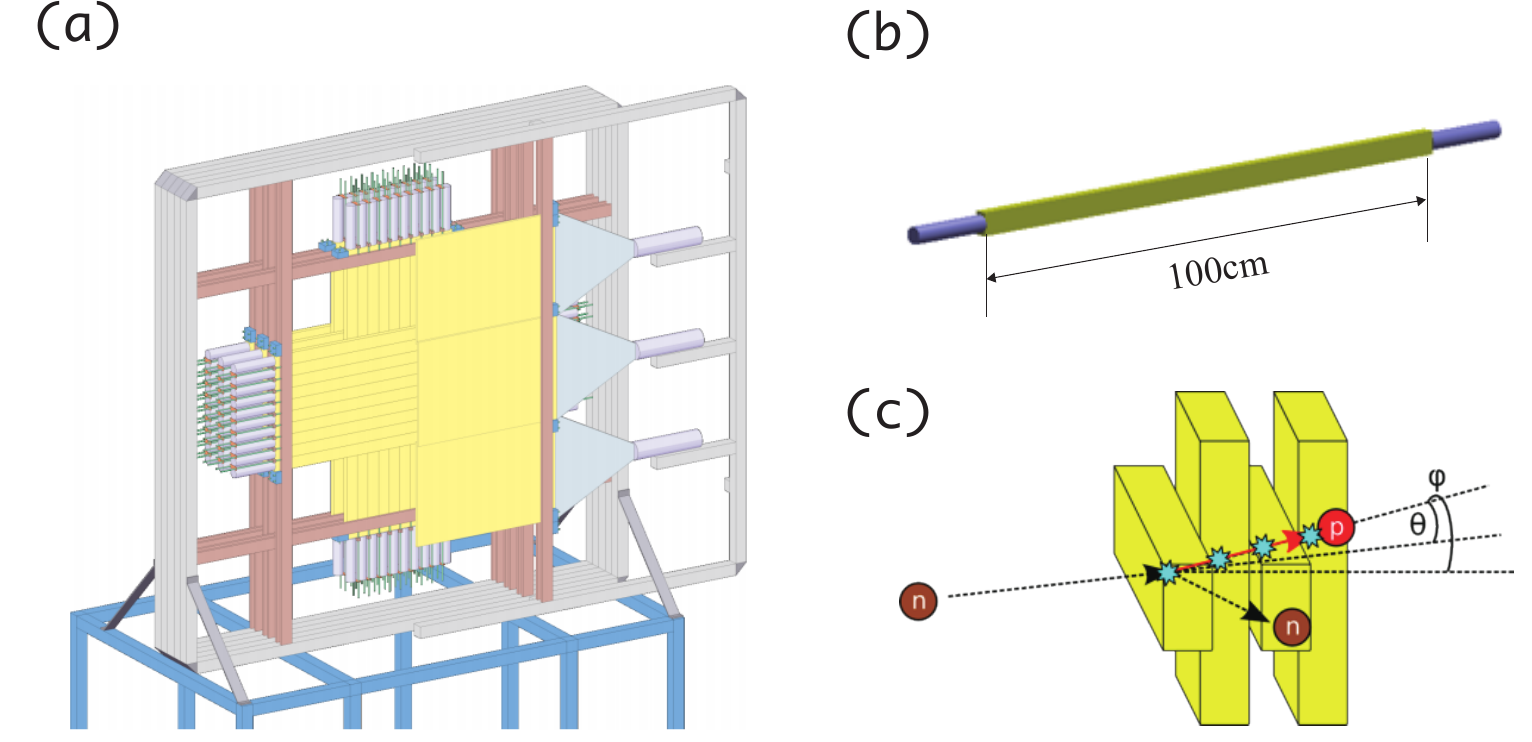}
   \caption{(a) One-wall configuration of the high-granularity neutron detector array HIME. One wall has five layers, each consisting of 10 plastic scintillator modules. We also install three VETO detectors with 1~cm thickness, half of which are shown. For two-neutron detection, we use a two-wall setting. (b) One scintillator module of HIME, which has dimensions of 4$\times$2$\times$100 cm$^3$. (c) Schematic concept of tracking of a recoil proton in the neutron detection}
   \label{fig:hime}
   \end{center}
\end{figure}

HIME is composed of 100 pieces of plastic scintillator modules, each of which is 
100~(H or V)~$\times$ ~ 4~(V or H)$\times$ ~ 2 ~ (D)~cm$^3$ in dimension 
and is coupled to two photo-multiplier tubes at both ends, as shown in Fig. \ref{fig:hime} (b). 
These are arranged in two wall configurations for two neutron-detection experiments, such as those for $^{26}$O.
One wall setting is schematically shown in Fig.\ref{fig:hime}~(a), which comprises 10~modules $\times$ 5 layers, and the direction of each layer is alternated by 90 degrees one by one.
The three Veto plastic scintillator bars with 1~cm thickness 
are also installed to veto charged particle backgrounds. 

Since the thickness of each HIME module is only 2~cm, most recoiled protons can penetrate into the next layer of the neutron detector at $\sim$200$-$300 MeV, which is the typical neutron energy at RIBF. Hence,  
the 3-densional hit position 
can be extracted using the identification of two or more bars that are fired, as shown in Fig.\ref{fig:hime}~(c). A good position resolution 
of $\sigma_x\sim 2$~cm and a timing resolution of $\sigma_t\sim 100$~ps are expected.
These features are essential for measuring the two-neutron angular correlation at very low relative energies, such as in the $ 2n$ decay of $^{26}$O.
 
%\subsubsection{Three-body decay of $^{6}$He(2$^+$)}

\subsection{Theoretical aspects of dineutron in unbound nuclei}
\label{sec:unboundtheo}

\begin{figure}[tb]
 \begin{center}
   \includegraphics[width=7.cm]{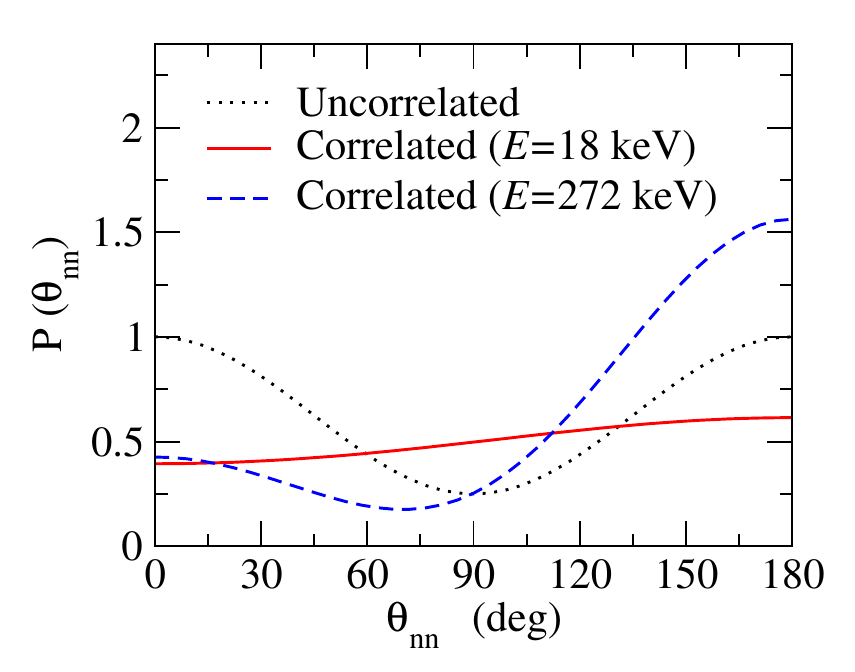}
   \caption{The angular correlation between the momenta of the emitted neutrons from the 
   unbound state of $^{26}$O~\cite{HAGI14,HAGI16}. The solid and the dashed lines show the results for the correlated 
   emissions with the energy of the unbound state $E$ of 18 and 272 keV, respectively, while the dotted line shows the 
   result for the uncorrelated emission
}
   \label{fig:26o-angular}
   \end{center}
\end{figure}

The angular correlation between the momenta of the emitted neutrons from an unbound state 
has been investigated with the three-body models.  
Fig. \ref{fig:26o-angular} shows the angular correlation for $^{26}$O studied in \cite{HAGI14,HAGI16}. 
The figure shows the results for two different resonance energies, $E=18$ keV (solid line) and 272 keV (dashed 
line). Both of these calculations show an asymmetric distribution, with an enhancement for the back-to-back emission 
$\theta_{nn}\sim\pi$. 
%\blue{\sout{Notice that the asymmetry for $E=272$ keV is somewhat larger than that for the empirical value \cite{KOND16}
%of $E=18$ keV.}} 
The asymmetric distribution is in contrast to the distribution for the uncorrelated case (the dotted line), which shows a completely symmetric behavior 
between the forward and the backward emissions. 

\begin{figure}[tb]
 \begin{center}
   \includegraphics[width=6.cm]{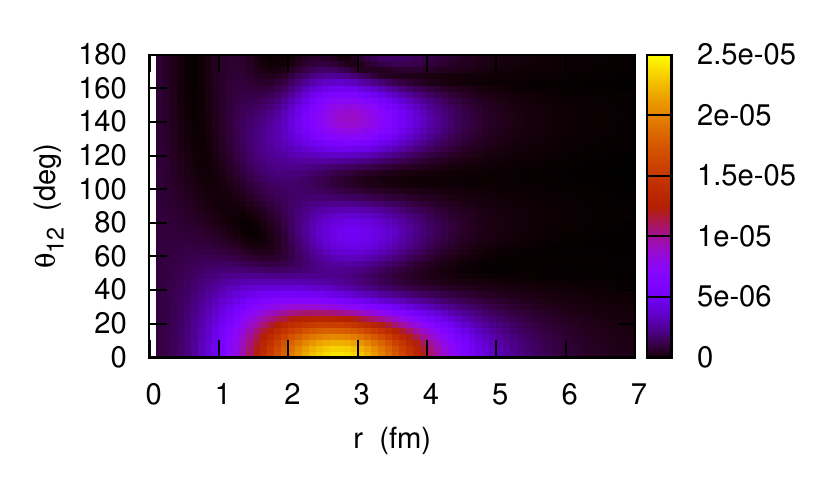}
      \includegraphics[width=6.cm]{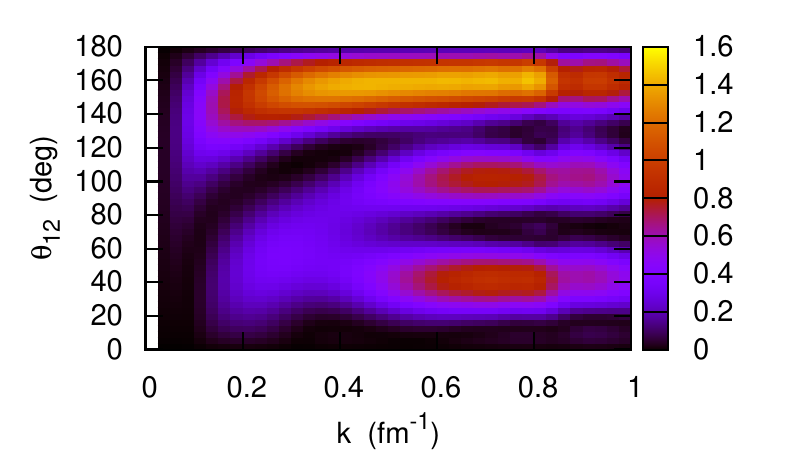}
   \caption{The two-particle density for the unbound state of $^{26}$O at $E=18$ keV 
   obtained with the three-body model of Refs. \cite{HAGI14,HAGI16}. The left and the right panels show the density 
   in the coordinate space ($|\Psi(r_1=r_2=r,\theta_{12})|^2$) and in the momentum space ($|\tilde{\Psi}(k_1=k_2=k,\theta_{12})|^2$), respectively. 
   For the latter, the weight factor of $8\pi^2k^4\sin\theta_{12}$ has been multiplied 
   }
   \label{fig:26o-density-rp}
   \end{center}
\end{figure}

%The asymmetry in the angular correlation is a direct consequence of the dineutron correlation (see the discussion in Sec.~\ref{sec:unboundexp}). 
The asymmetry in the angular correlation should be related to the dineutron correlation (see the discussion in Sec.~\ref{sec:unboundexp}) if the two-neutron emission process reflects the initial wave function before the decay.
If one takes the Fourier transform of Eq. (\ref{eq:3-body-oddeven}), one finds
\begin{equation}
\tilde{\Psi}(\vec{k}_1,\vec{k}_2)
\equiv
\int d\vec{r}_1\vec{r}_2\,e^{i\vec{k}_1\cdot\vec{r}_1}e^{i\vec{k}_2\cdot\vec{r}_2}
\Psi(\vec{r}_1,\vec{r}_2)
=\alpha \tilde{\Psi}_{\rm ee}(\vec{k}_1,\vec{k}_2)-\beta \tilde{\Psi}_{\rm oo}(\vec{k}_1,\vec{k}_2), 
\end{equation}
where $\tilde{\Psi}_{\rm ee}$ and $\tilde{\Psi}_{\rm oo}$ are the Fourier transforms of 
$\Psi_{\rm ee}$ and $\Psi_{\rm oo}$, respectively. 
Notice that the second term in the last equation has the opposite sign to that in the wave function in the 
coordinate space, Eq. (\ref{eq:3-body-oddeven}). This is understood from the formula for the partial wave 
decomposition, 
\begin{equation}
e^{i\vec{k}\cdot\vec{r}}=4\pi\sum_l i^lj_l(kr)Y_l^*(\hat{\vec{r}})\cdot Y_l(\hat{\vec{k}}), 
\end{equation}
which indicates that the Fourier transform of 
$[\phi_{njl}(\vec{r}_1)\phi_{njl}(\vec{r}_2)]^{(00)}$ involves the factor $i^{2l}=(-1)^l$. 
That is, if the coefficients $\alpha$ and $\beta$ in Eq. (\ref{eq:3-body-oddeven}) are such that 
the dineutron component with $\vec{r}_1\sim\vec{r}_2$ is enhanced over the cigar-like component 
with $\vec{r}_1\sim -\vec{r}_2$, the opposite happens in the momentum space and 
the component with $\vec{k}_1\sim -\vec{k}_2$ is more enhanced than the component with  $\vec{k}_1\sim \vec{k}_2$. 
This is actually seen in the two-particle density for $^{26}$O in the momentum space obtained with the three-body model of Refs. \cite{HAGI14,HAGI16} (see Fig. \ref{fig:26o-density-rp}). 

%\blue{
In Fig. \ref{fig:26o-angular}, the asymmetry for $E=272$ keV is somewhat larger than that for the empirical value \cite{KOND16}
of $E=18$ keV.
For $E=18$ keV, the neutron emission from the $p$-wave is largely suppressed due to the centrifugal barrier, 
and the admixture between the $s$-wave and the $p$-wave components is hindered in 
the final wave function, leading to the smaller asymmetry in the angular distribution. For $E=272$ keV, the admixture is larger, and thus the asymmetry in the angular distribution is significantly enhanced. 
%}

\begin{figure}[tb]
 \begin{center}
   \includegraphics[width=10.cm]{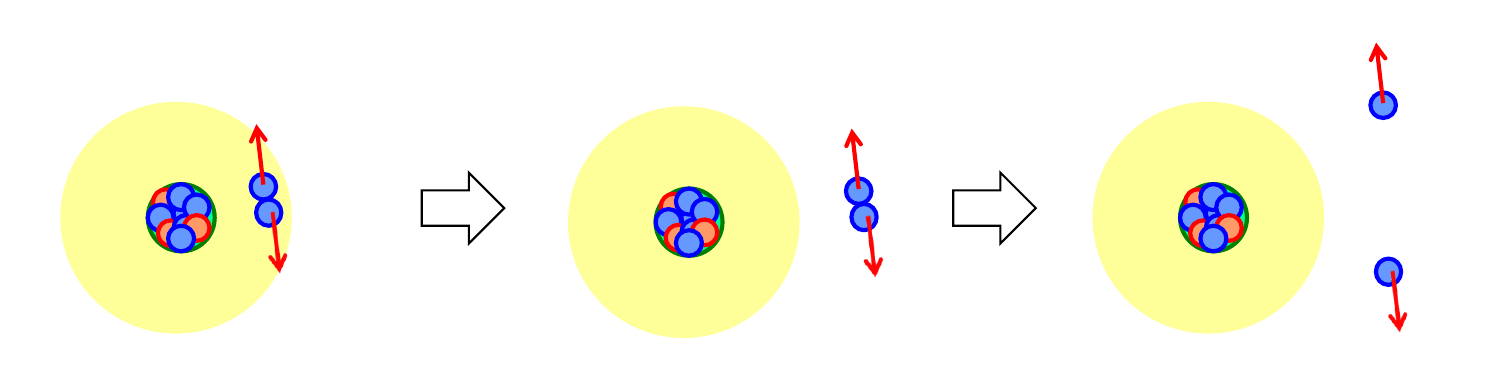}
   \caption{A schematic illustration for a role of dineutron correlation in a two-neutron emission from an 
   unbound nucleus. Two neutrons are first confined within a nucleus with angular momenta in the opposite 
   direction (the left figure). They are then emitted from the core nucleus 
   to a similar direction reflecting the dineutron 
   correlation in the coordinate space (the middle figure). After the two neutrons are outside the core 
   nucleus, they fly apart in the opposite direction reflecting the dineutron correlation in the momentum 
   space (the right figure)
   }
   \label{fig:2ndecay-schematic}
   \end{center}
\end{figure}

The opposite asymmetry between the coordinate and the momentum spaces leads to an interesting 
consequence for a two-neutron emission, as is shown in Fig.~\ref{fig:2ndecay-schematic}. 
Two neutrons are first confined within a nucleus with a spatially compact shape but with a large relative 
momentum. At the early stage of two-neutron emission, those two neutrons are emitted in a similar direction, 
reflecting the spatially compact configuration. However, once the two neutrons are outside a nucleus, there 
is no mechanism to spatially confine them, and they fly apart in the opposite direction, reflecting the large 
relative angular momentum. This scenario has actually been confirmed in the time-dependent three-body model calculations 
for two-proton decays of 
unbound nuclei \cite{OISH14,WANG21}, which are similar phenomena to two-neutron emissions from unbound nuclei. 

\section{Tetraneutron}\label{sec:tetra}
Here we review recent tetraneutron experiments, with some emphasis on dineutron aspects.

\subsection{Experiments on tetraneutrons}
Since the 1960s, tetraneutrons have been searched for by many experiments using fission, pion-induced double charge exchange reaction, and multi-neutron transfer (see Ref.~\cite{MARQ21}). 
Most studies report no observation of tetraneutrons, while a few experiments report positive signals with very limited statistics.
One of the positive results is given by measuring a multi-neutron cluster produced in the breakup reaction $^{14}$Be$\rightarrow^{10}$Be+$4n$ at 35~MeV/nucleon in 2002 at GANIL~\cite{MARQ02}.
The measurement with a neutron detector array (DEMON) reports six candidate events of a tetraneutron cluster.
It can be interpreted as bound four neutrons or a low-energy 
four-neutron resonance.

\begin{figure}[tb]
 \begin{center}
   \includegraphics[width=7cm]{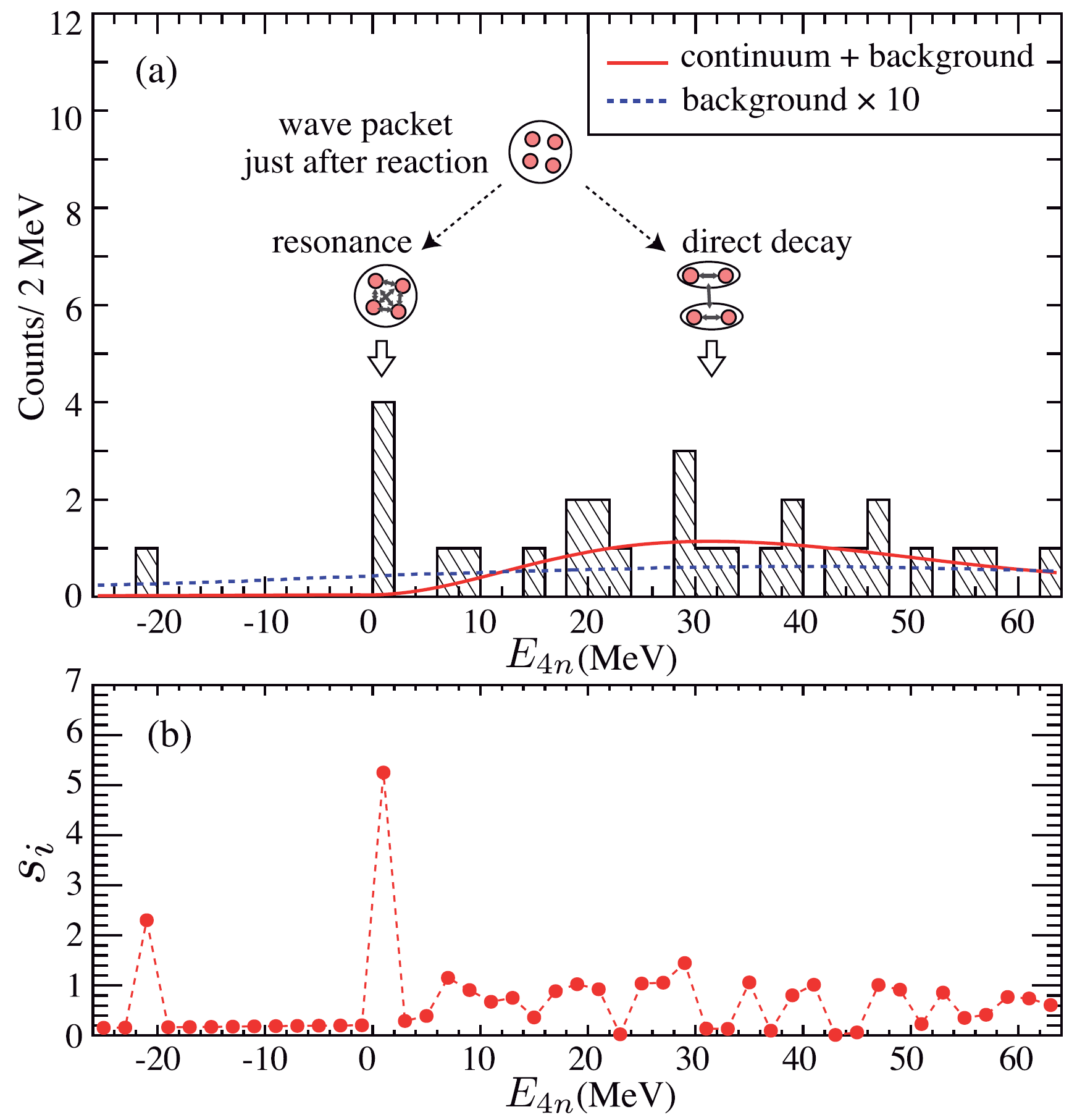}
   \caption{(a) Missing mass spectrum of four-neutron system observed in the $^4$He($^8$He,~$^8$Be) reaction. (b) Evaluation of the goodness of fit using the likelihood ratio test. The figure is adopted from Ref.~\cite{KISA16}}
   \label{fig:4n-kisamori}
   \end{center}
\end{figure}

The last decade has seen great progress in the experimental study of the tetraneutron.
A missing mass measurement~\cite{KISA16} in the double charge exchange reaction $^4$He($^8$He,~$^8$Be)$^4n$ at 186~MeV/u
was performed using the SHARAQ spectrometer~\cite{UESA12} at RIBF.
%The reaction is suitable for producing the tetraneutron system with almost recoilless conditions.
Owing to the large positive Q value of the ($^8$He,~$^8$Be) reaction, an almost recoilless condition is achieved for populating the weakly interacting system efficiently.
Four candidate events were observed in the region 0--2~MeV in the four-neutron energy spectrum (Fig.~\ref{fig:4n-kisamori}), consistent with a tetraneutron resonance.
The significance level is 4.9$\sigma$ in comparison with the continuum that assumes no resonant state.
The energy of 0.83$\pm$0.65(stat)$\pm$1.25(syst)~MeV above the four-neutron decay threshold and the upper-limit width of 2.6~MeV (FWHM) were obtained.

\begin{figure}[tb]
 \begin{center}
   \includegraphics[width=7cm]{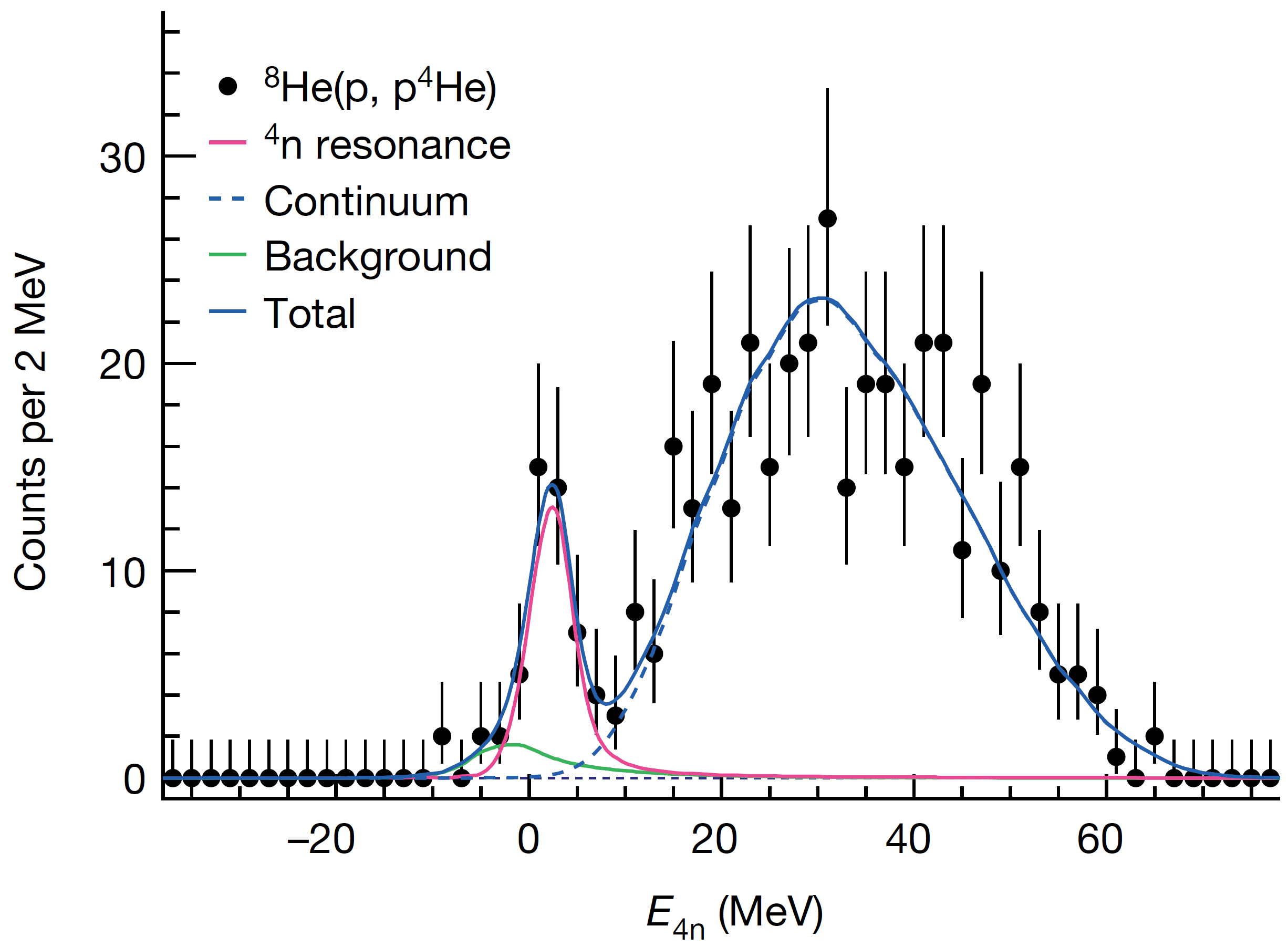}
   \caption{Missing mass spectrum of four-neutron system observed in the $^{8}$He($p$,~$p^4$He) reaction.
     The figure is reproduced from Ref~.\cite{DUER22} under CC BY 4.0 (\url{http://creativecommons.org/licenses/by/4.0/})}
   \label{fig:4n-duer}
   \end{center}
\end{figure}

This is followed by a missing mass measurement with higher statistics using the $^8$He+$p$ reaction at 156~AMeV~\cite{DUER22}, where the quasi-elastic knockout of $\alpha$ in $^8$He produces a tetraneutron system.
The selection of kinematics at large momentum transfer between the proton and the $\alpha$-particle minimizes the final-state interactions among the outgoing four neutrons and the charged particles.
The large acceptance spectrometer SAMURAI~\cite{KOBA13} enabled the simultaneous detection of the outgoing $\alpha$ and proton.
A missing mass spectrum (Fig.~\ref{fig:4n-duer}) shows a narrow, pronounced peak that includes about 50 counts at the energy of 2.37$\pm$0.38(stat)$\pm$0.44(syst)~MeV above the four-neutron decay threshold with a width of 1.75$\pm$0.22(stat)$\pm$0.30(syst)~MeV, which is consistent with the result of the SHARAQ experiment~\cite{KISA16}.

Another missing mass measurement~\cite{FAES22} also reports the positive signal
using a different reaction, $^7$Li($^7$Li,$^{10}$C) at 46~MeV.
A peak was observed at 2.93(16)~MeV at the excitation energy of the $^{10}$C+4$n$ system with significance of $\sim$3$\sigma$.
The results give two possible interpretations: the tetraneutron is unbound by 2.93~MeV with an extraordinarily small width $\Gamma < 0.24$ or the $^{10}$C is in the first excited state and the tetraneutron is bound with a binding energy of 0.42(16)~MeV.

Theoretically, there are positive and negative results about the existence of the tetraneutron state. Shirokov et al. applied various ab initio methods using the JISP16 realistic $NN$ interactions to extract the resonance energy of $E_{4n}=0.8$ MeV with $\Gamma=1.4$~MeV~\cite{SHIR16}. Gandolfi et al. used quantum Monte Carlo calculations of few-neutron systems confined in external potentials based on local chiral interactions at the N$^2$LO chiral effective field theory, and assumed that the extrapolation to zero external potential depth provides an estimate of the $4n$ state~\cite{GAND17}. They find that the $4n$ system exhibits a resonance at $E_{4n} \sim 2$ MeV. These predictions are in line with the results of the two experiments at RIKEN~\cite{KISA16,DUER22}.
The calculation in Ref.~\cite{GAND17} also finds 
that the energy of $3n$ is lower than that of $4n$, while such a $3n$ resonance is not observed in the recent experiment~\cite{MIKI24}.

%, which is against the recent non-observation of $3n$~\cite{MIKI24}.

On the other hand, there are objections to the observation of the tetraneutron.
Deltuva studied the tetraneutron using exact continuum equations for transition operators~\cite{DELT18}. Resonance behavior is found for strongly enhanced interactions, but not for the physical strength, negating an observable tetraneutron resonance.
This study also suggested an enhanced behavior in low $E_{4n}$, and conjectured that such a low-energy enhancement might have been observed in the tetraneutron experiment.
%Deltuva and Lazauskas used a rigorous treatment of the four-particle continuum and investigated the $4n$ system with artificially enhanced interaction~\cite{DELT19}. This study finds that the tetraneutron evolves not into a resonance but into a virtual state, which decays into two dineutrons. The result shows that no $4n$ resonance should be observed.
 Higgins et al. studied the $4n$ system using the adiabatic hyperspherical framework to understand low-energy $4n$ states in terms of adiabatic potential energy curves~\cite{HIGG20}. The result shows no sign of a low-energy resonance. Their calculation showed a low-energy enhancement of the density of states, which may be relevant to the experimental observations.

Lazauskas et al.\cite{LAZA23} theoretically examined the result of the $^{8}$He($p,p\alpha$)$4n$ experiment~\cite{DUER22}.
This study investigated the reaction model that describes a fast removal of $\alpha$ from $^8$He, and could show that the peak observed in Ref.~\cite{DUER22}
can be explained by the dineutron-dineutron correlation in the initial state of $^{8}$He.
This shows that tetraneutron studies require a full understanding of the initial state and reaction dynamics. The association with the dineutron-dineutron correlation is an interesting topic, as such a state may show a BEC-like structure composed of two bosons. More experimental and theoretical studies on $4n$ and the initial state $^8$He are needed in the near future.
We also note that we should understand why the double charge-exchange reaction~\cite{KISA16} and ($p,p\alpha$) reaction~\cite{DUER22} provide consistent results at $E_{4n}\sim 1-2$~MeV.

\subsection{Experiments on 4n system in neutron-rich nuclei}
\subsubsection{$^{28}$O}
The $^{28}$O nucleus has attracted much attention as a candidate for a doubly magic nucleus with $Z$=8 and $N$=20.
However, only its unbound nature against the four-neutron emission has been experimentally established~\cite {TARA97,SAKU99}.
%Since the contribution of the neutron $pf$ shell across the $N$=20 shell gap is large in the neighboring region, called the island of inversion, it is a question whether $^{28}$O is a doubly magic nucleus or the island of inversion is extended to oxygen.
In a neighboring region, known as the island of inversion, the contribution of the neutron $pf$ shell plays a significant role across the $N$=20 shell gap (see e.g., Ref.~\cite{OTSU20}).
The island of inversion has been shown to extend to $^{29}$F, which exhibits a low-lying excited state~\cite{DOOR17} and a two neutron halo structure attributed to a neutron $p_{3/2}$ orbital~\cite{BAGC20}.
It is a question whether $^{28}$O is a doubly magic nucleus or lies within the island of inversion extending down to oxygen.

\begin{figure}[tb]
 \begin{center}
   \includegraphics[width=\textwidth]{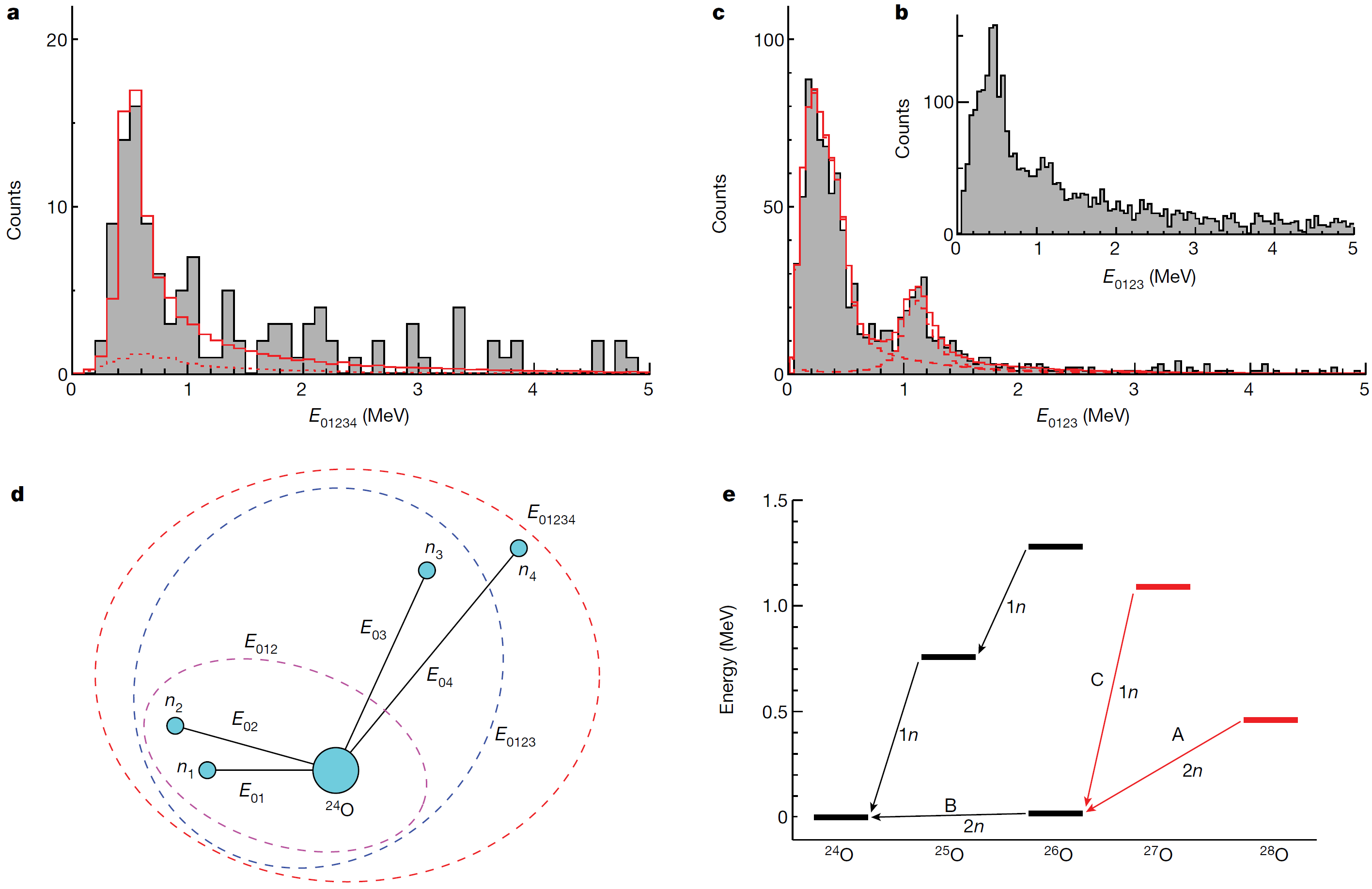}
   \caption{a) Spectrum of five-body ($^{24}$O+4$n$) decay energy $E_{01234}$.
     b)c) Spectra of four-body ($^{24}$O+3$n$) decay energy without and with tagging the decay of the $^{26}$O ground state ($E_{012}<0.08$~MeV).
     d) Definition of decay energies.
     e) Decay scheme of the unbound oxygen isotopes. The figure is reproduced from Ref.\cite{KOND23} under CC BY 4.0 (\url{http://creativecommons.org/licenses/by/4.0/})}
   \label{fig:28o-spectrum}
   \end{center}
\end{figure}

The first observation of $^{28}$O~\cite{KOND23} was made using the SAMURAI spectrometer~\cite{KOBA13} at RIBF.
It is produced by one-proton removal from a 235-MeV/nucleon beam of $^{29}$F.
The invariant mass of the neutron unbound nucleus is obtained from the measured momentum vectors of decay particles $^{24}$O + 4$n$.
Special care was taken to treat multiple interactions of single neutrons in neutron detectors, called neutron crosstalk, as it gives false two or more neutron events~\cite{NAKA16,KOND20}.
The $^{28}$O ground state resonance has been observed at 0.46$^{+0.05}_{-0.04}$(stat)$\pm 0.02$(syst)~MeV in the decay energy spectrum of $^{24}$O+4$n$ (Fig.~\ref{fig:28o-spectrum}a).
In addition, a resonance of $^{27}$O has also been identified for the first time at 1.09$\pm 0.04$(stat)$\pm 0.02$(syst)~MeV in the decay energy spectrum of $^{24}$O+3$n$~(Fig.~\ref{fig:28o-spectrum}b).
The obtained spectroscopic factor 0.48$^{+0.05}_{-0.06}$(stat)$\pm 0.05$(syst) from the measured cross section of the one-proton removal from $^{29}$F indicates that the neutron configuration of $^{28}$O has a large overlap with the island-of-inversion nucleus $^{29}$F.
This leads to the conclusion that the $N=20$ shell gap disappears in $^{28}$O.

\begin{figure}[tb]
 \begin{center}
   \includegraphics[width=\textwidth]{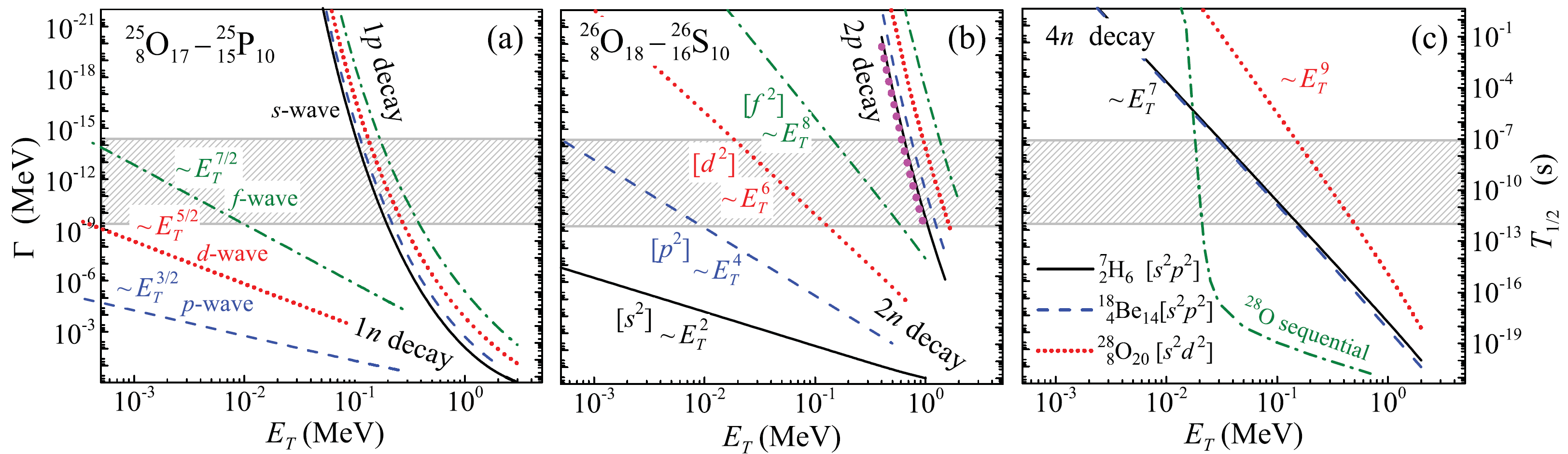} \caption{Estimated decay widths as a function of decay energy for (a) one-nucleon, (b) two-nucleon, (c) four-neutron emission. The figure is adopted from Ref.~\cite{GRIG11}}
   \label{fig:width-grigorenko}
   \end{center}
\end{figure}

Investigation of the decay energies of sub-systems shows that both $^{28}$O and $^{27}$O sequentially decay through the ground state of $^{26}$O, which has a small decay energy of 18~keV (see also Sec.~\ref{sec:o26})~\cite{KOND16}.
%The sequential ($2n$)-($2n$) of $^{28}$O is consistent with theoretical predictions~\cite{GRIG11}, showing that true four-neutron emission is strongly hindered compared to two-neutron emission (Fig.~\ref{fig:width-grigorenko}) due to effective few-body centrifugal barriers, which increase rapidly with the number of particles emitted.
The sequential ($2n$)-($2n$) emission of $^{28}$O is consistent with theoretical predictions~\cite{GRIG11}. As shown in Fig.~\ref{fig:width-grigorenko}(c), the width of the true four-neutron emission is much narrower (red dotted line) than that for two-neutron emission (green dot-dashed line) due to effective few-body centrifugal barriers, which increase rapidly with the number of particles emitted.
Taking into account the doubly magic nature of $^{24}$O~\cite{OZAW00,OTSU01,HOFF09,KANU09,TSHO12} and the predicted dineutron correlation in $^{26}$O~\cite{HAGI14},
$^{28}$O may be viewed as a system with two dineutron clusters.
%As the partial decay energies of $^{28}$O$\to^{26}$O + 2$n$ and $^{26}$O$\to^{24}$O + 2$n$ are well separated in energy, the angular correlation of two dineutron clusters is expected to reflect the correlations of the two dineutron clusters.
As the partial decay energies of $^{28}$O$\to^{26}$O + 2$n$ and $^{26}$O$\to^{24}$O + 2$n$ are well separated in energy, emissions of the two possible dineutron clusters can be identified experimentally. This enables the study of the angular correlation of the two dineutron cluster emissions, which is expected to reflect their spatial correlations.
A next-generation neutron detector array with high resolution and high efficiency is needed to realize the measurement.

As a dineutron cluster is developed if different parity configurations are mixed, knowledge on neutron shell structure and occupancy of single particle orbitals are essential for understanding of dineutron cluster(s). One of the key nuclei is $^{27}$O, as its low-lying states and spin-parities should reflect the neutron shell structure, while they are still unclear.
A recent study~\cite{REDP24} discusses the possible spin and parity of the $^{27}$O resonance and the structural change between neutron-rich oxygen and fluorine/neon isotopes from measured cross sections for proton removal reactions.
In the two-proton removal from the 3/2$^-$ of $^{29}$Ne,
only a few events were observed for the decay path of $^{27}$O via the $^{26}$O ground state,
which is much smaller than expected from the shell-model calculations assuming 3/2$^-$ for $^{27}$O.
The other possible scenarios, $3/2^+$ and $7/2^-$, also require a structural change of $^{27}$O from $^{29}$Ne.
In contrast, the measured cross section for populating the ground state of $^{26}$O from $^{27}$F is consistent with a large spectroscopic factor of 0.84, as predicted by shell-model calculations.
The structural change between $^{25}$F and $^{24}$O has also been reported~\cite{TANG20,CRAW22}, where a measured spectroscopic factor of 0.36(13) for populating the ground state of $^{24}$O conflicts with the shell model calculations, 
while a large overlap is observed in the one-proton removal from $^{29}$F to $^{28}$O~\cite{KOND23}.
%\red{\sout{Further experimental and theoretical studies are desired to understand the evolution of structural change in the region.
%}}

\begin{figure}[htb]
 \begin{center}
   \includegraphics[width=70.mm]{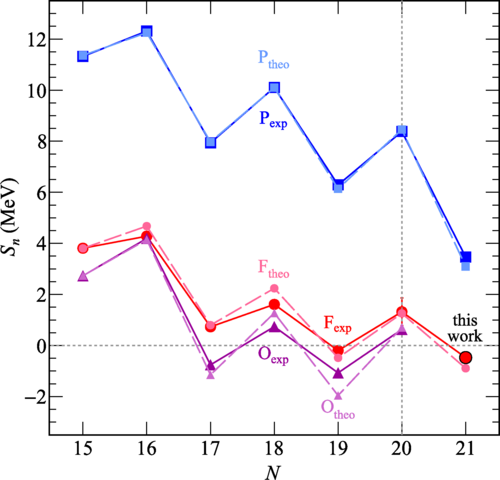} \caption{
   Systematics of one-neutron separation energies of neutron-rich oxygen and fluorine isotopes, which are compared to the phosphorus isotopic chain. The experimental data points are connected by solid lines, while the shell model calculations are connected by dashed lines. The result for $^{30}$F is marked as ``this work". 
   The disappearance of $N=20$ is seen for fluorine isotopes.
   The figure is reproduced from Ref.~\cite{KAHL24} under the Creative Commons Attribution 4.0 International (CC BY 4.0) license }
   \label{fig:30F}
   \end{center}
\end{figure}

The neighboring nucleus, $^{30}$F ($Z=9, N=21$), was recently
observed using the $^{31}$Ne($p,2p$)$^{30}$F reaction at SAMURAI at RIBF, where the ground-state mass of $^{30}$F was determined: $\sn= -472\pm58({\rm stat})\pm 33 ({\rm sys})$ keV~\cite{KAHL24}. 
Thus, $^{30}$F is barely unbound with respect to one-neutron emission. 
This result, as well as those for $^{27,28}$O, are included in the systematics of one-neutron separation energy shown in Fig.~\ref{fig:30F}.
Compared with the systematics for phosphorus isotopes, it is clear that the $N=20$ shell gap disappears for fluorine isotopes. 
The nearly identical oscillation pattern from $N=17$ to $N=20$ for oxygen and fluorine isotopes, compared to the large-scale shell model calculations with SDPF-U-MIX20 interactions, shows that in these nuclei
the neutron orbitals $(\nu 0d_{3/2})^2$, $(\nu 0f_{7/2})^2$, $(\nu 1p_{3/2})^2$, and $(\nu 1p_{1/2})^2$ lie close in energy. On the other hand, spherical nature persists due to the large proton gap at $Z=8$ in $^{28}$O, which enhances the correlation for neutron pairs coupled to $J=0$. The degeneracy of neutron orbitals in a spherical mean field implies that
neutron pairs near the Fermi surface can effectively hop between different orbitals through pair scattering. This process provides the microscopic mechanism for pairing correlations leading to superfluidity~\cite{BRIBRO05}.
%Since these neutron orbitals are close to each other a pair of neutrons can be scattered with each other to hopping to other 
%The closeness of these neutron orbitals, combined with the fact that the , shows the nature of superfluidity in this nucleus.
The shell-model calculation shows
that $^{28}$O has 97\% of pairs with $J=0$, corresponding to seniority zero,
while it has 50\% closed configuration and 47\% of 2$\hbar\omega$ configurations (mostly $2p2h$ excitations from $\nu 0d_{3/2}$ to $\nu 1p_{3/2}$). A similar trend is also obtained for $^{29}$F.
The result shows that pairing correlations indeed play critical roles
in such extremely neutron-rich nuclei.
Further experimental and theoretical studies are called for to understand the evolution of structural changes in this region.

\subsubsection{$^{8}$He}\label{sec:8He}

$^8$He is an intriguing, extremely neutron-rich nucleus, as it is a neutron skin nucleus that comprises the inert $\alpha$ core surrounded by the four valence neutrons~\cite{TANI92}, having the highest ratio $N/Z$ ($N / Z = $3) among all the bound nuclei.
Unlike most neutron drip-line nuclei, the two-neutron separation energy ($S_{2n}=2.1250(1)$~MeV) is
larger than that of the less neutron-rich isotope $^{6}$He ($\snn=0.97546(5)$~MeV~\cite{AME2020}). As such, $^8$He may have $4n$-cluster or double dineutron structures.
As discussed above, the $4n$ experiment used the reaction 
$^8$He($p,p\alpha$)~\cite{DUER22},
and the theoretical work~\cite{LAZA23} suggests the effect of the dineutron-dineutron structure in $^8$He for the low-lying energy spectrum of the four neutrons. Hence, the understanding the structure of $^8$He is also important in this sense.

The $0^+_2$ excited state of $^8$He has been predicted to be a candidate for a dineutron condensated state by theoretical studies using the antisymmetrized molecular dynamics (AMD) method~\cite{KANA07} and the $^2n$ cluster model~\cite{KOBA13b,NAKAG25}.
Experimentally, the excited $0^+$ state was identified at 4.5~MeV in the decay energy spectrum reconstructed from the measured momentum vectors of decay particles, $^6$He and two neutrons, for inelastic excitation by (CH$_2$)$_n$ and carbon targets at 82.3~MeV/nucleon~\cite{YANG23}.
A large isoscalar monopole transition strength $M$(IS0)=$11^{+1.8}_{-2.0}$~fm$^2$ deduced from the measured cross section is consistent with the predicted value of 9.0~fm$^2$, associated with a condensate-like cluster structure~\cite{KOBA13b}. 
Recently, Nakagawa and Kanada En'yo applied a microscopic cluster model with the generator coordinate method for the $\alpha$+$^2n$+$^2n$ cluster structures and the breaking of $^2n$ clusters~\cite{NAKAG25}. They found that both the ground and excited 0$^+$ states ($^8$He($0_1^+$), $^8$He($0_2^+$)) have a dominant (84\%, 93\%) $\alpha$+$^2n$+$^2n$ cluster structure (
.~\cite{KANA07,HAGI08,ITAG08,KOBA13b,YAMA23,NAKAG25} for related works). 
In addition, this theory could explain reasonably well the experimental isoscaler monopole transition shown above.

%It is interesting to note that 

\section{Summary and Perspective on dineutron correlations}\label{sec:summary}

Recent experimental and theoretical developments on dineutron correlations in neutron-rich nuclei have been discussed. The compact spatially correlated dineutron is expected to occur on the low-density part of nuclei and nuclear matter. Valence neutrons in neutron-halo nuclei and barely unbound states are candidate sites for dineutrons. Dineutron correlations are
%{\sout{has an analogy to} 
closely related to the BCS-BEC crossover observed in general fermionic systems. The pairing correlations in the BCS-BEC crossover appear near the unitary limit, while the $nn$ scattering length is large in magnitude ($|a|=18.9(4)$~fm). 
The studies of dineutron phenomena may thus be relevant to the physics of ultra-cold atoms~\cite{HORI25} and to diquarks~\cite{HOSA25} in hadron physics.
We also note that the dineutron on the surface of a nucleus appears in the threshold to two-neutron decay, which has a common feature with $\alpha$ clusters and hadronic molecules~\cite{NAKA25,HOSA25}: Clusters develop near the threshold. 

For halo nuclei, dineutrons of the Borromean two-neutron halo nuclei have been discussed. Coulomb breakup and charge radii measurements 
have played central roles in establishing the compact dineutron picture in halo nuclei, such as $^6$He and $^{11}$Li.
In these methods, one can derive the shift of the core center relative to the c.m. of the whole system, thereby yielding the geometric mean opening angle between the two valence neutrons. More recently, the experimental approach using proton-induced quasi-free $(p,pn)$ scattering on two-neutron halo nuclei has been developed. The case of $^{11}$Li showed that the compact dineutron appears just on the surface of the core~\cite{KUBO20}. The dineutron disappears not only within the core but also far from it. 
Recently, Corsi et al.~\cite{CORS23} extended $(p,pn)$ studies to $^{14}$Be and $^{17}$B, where dineutron properties are also confirmed, and as such, the compact dineutron property can be universal for any two-neutron halo nuclei. 

Theoretically, three-body model calculations have been developed to understand the dineutron nature of two-neutron halo nuclei. 
This model consists of a core nucleus and two valence neutrons. Based on this model, the dineutron correlation is caused by 
an admixture of 
two-particle configurations involving even-parity and 
odd-parity single-particle states. 
The same model can be applied to both the ground state and the excited states simultaneously. For instance, the total $\eone$ strength 
is related to the ground state properties of a Borromean nucleus through the cluster sum rule. Using this method, the opening angle 
between the valence neutrons in $^{11}$Li and $^6$He was extracted. The resultant values were significantly smaller than 90 degrees 
for both nuclei, indicating the presence of the dineutron correlation in these nuclei. 

For unbound nuclei, recently identified neutron-rich states include two-neutron emitters such as $^{16}$Be and $^{26}$O, and four-neutron emitters such as $^{28}$O. For $^{16}$Be, a hint of dineutron correlation in the ground state has  recently been suggested by comparisons with three-body calculations that incorporate decay dynamics. For $^{26}$O, the angular correlation is a promising method, but the angular resolution has not been sufficient for such a measurement. 
The recent development of the high-granularity neutron detector array, HIME, at RIBF, RIKEN, would be necessary for such a study. At RIKEN, we are currently developing an even more novel neutron detector array, called NEOLITH, that combines traditional plastic scintillator converters with tracking detectors (drift chambers) for recoil protons. As such, we expect the position resolution 
to be $\sim$~mm, and the angular resolution in the laboratory frame to be $\sim$~0.1~mrad. The details of NEOLITH will be published elsewhere. 
In the coming years, we expect that the properties of dineutron correlation in such barely unbound two-neutron emitters will be experimentally clarified.

Theoretically, the three-body model has been applied to the two-neutron emission decay of unbound nuclei. 
It has been shown that the back-to-back emission of two neutrons is enhanced, reflecting the dineutron correlation in the 
ground state. 

Recent observations of the $^4n$ candidate states at RIBF, RIKEN, have triggered studies of pure-neutron nuclei (multi-neutron systems) both experimentally and theoretically. An objection to this observation was that the dynamics and initial state play a crucial role, and Lazauskas claimed that the peak observed in the experiment at RIKEN~\cite{DUER22} can be understood in terms of the dineutron-dineutron correlation in the initial state of $^8$He~\cite{LAZA23}. Indeed, understanding the dineutron-dineutron or tetraneutron cluster structure in $^{8}$He is essential for interpreting the $4n$ spectrum shown in Fig.~\ref{fig:4n-duer}. 
Recently, R. Garrido and A.S. Jensen~\cite{GARR25} 
have discussed the $\alpha$+$^2n$+$^2n$ cluster structure in $^8$He by varying the $^2n$-$^2n$ interactions. They find pronounced differences in the $4n$ spectrum in the $^8$He$(p,p\alpha)$~\cite{DUER22}, depending on the $^2n$-$^2n$ interactions. 
%Hence, the $4n$ spectrum can serve as a probe of the dineutron-dineutron configuration and the corresponding molecular-like correlation in $^{8}$He. 
As mentioned in Sec.~\ref{sec:8He}, Nakagawa and Kanada En'yo showed more microscopic calculation that the $^8$He ground state has predominantly $\alpha$+$^2n$+$^2n$ configuration~\cite{NAKAG25}.
Such an approach, understanding the initial cluster configuration involving dineutron correlations, may provide a key to understanding 
the multi-neutron systems in future experiments with higher energy resolution. 
We also note the importance of microscopic structure theories to be combined with the reaction theories to relate the observables to the dineutron (or multi-neutron) clusters.

%In fact, the $4n$ cluster states are expected in $^8$He.
%We have discussed that ...

The heaviest oxygen isotope, $^{28}$O, observed recently at RIBF, RIKEN~\cite{KOND20}, is also a possible site for the $4n$ correlation or dineutron-dineutron correlation.
The dineutron–dineutron correlation suggests the possible emergence of a Bose–Einstein condensed state if the dineutron is treated as a bosonic degree of freedom. In the future, multi-dineutron systems should be investigated to clarify the nature of this novel bosonic state.

During the preparation of this review article, a new result on the Coulomb breakup of $^{8}$He has been published~\cite{DUER26}, in which the $^4$He+$4n$ channel is treated for the first time in addition to the $^6$He+$2n$ channel.  We also note a new result on the spectroscopy of $^{7}$He via neutron removal of $^8$He, where three neutrons were measured in coincidence with $^4$He~\cite{HUAN26}. Such multiple-neutron measurements are expected to clarify the properties of mutli-neutron clusters that emerge near and beyond the neutron drip line.

For these studies, a large-scale neutron detector array, such as NEBULA~\cite{NAKA16,KOND20}, NEBULA-plus at RIKEN, MoNA~\cite{BAUM05} at FRIB, and NeuLAND~\cite{BORE21} at FAIR-GSI, will be crucial. 
We also note that, to achieve higher energy and angular resolutions, as well as multi-neutron detection capabilities, a new-generation neutron detector array, such as HIME and NEOLITH, may provide key to future experiments.

\bmhead{Acknowledgments}

The present work was supported in part by JSPS KAKENHI Grant Nos. JP16H02179, JP18H05404, JP21H04465, JP21H00006, 
and JP23K03414. The authors thank Alexander Volya, Jes\'us Casal, and Yuki Kubota for fruitful and insightful discussions.

%%===========================================================================================%%
%% If you are submitting to one of the Nature Portfolio journals, using the eJP submission   %%
%% system, please include the references within the manuscript file itself. You may do this  %%
%% by copying the reference list from your .bbl file, paste it into the main manuscript .tex %%
%% file, and delete the associated \verb+\bibliography+ commands.                            %%
%%===========================================================================================%%

\bibliography{multi_neutron_epj}% common bib file
%% if required, the content of .bbl file can be included here once bbl is generated
%%\input sn-article.bbl

\end{document}